\documentclass[reprint, superscriptaddress, secnumarabic, amssymb, nobibnotes, aps, pra]{revtex4-1}

\usepackage[utf8]{inputenc}
\usepackage{graphicx}
\usepackage{subfiles}
\usepackage{epstopdf}
\usepackage{parskip}
\usepackage[T1]{fontenc}

\usepackage{amsbsy}
\usepackage{gensymb}

\usepackage[T1]{fontenc}

\usepackage{amsmath}
\usepackage{booktabs}
\usepackage{amssymb}
\usepackage{bbm}
\usepackage{physics}
\usepackage{braket}
\usepackage{xcolor}
\usepackage{ wasysym }
\usepackage{float}
\allowdisplaybreaks
\usepackage{graphicx}
\usepackage[colorlinks=true]{hyperref}  
\hypersetup{
    bookmarks=true,         
    unicode=false,          
    pdftoolbar=true,        
    pdfmenubar=true,        
    pdffitwindow=false,     
    pdfstartview={FitH},    
    pdftitle={Investigating Non-trivial Topology from Structural Transition Perspective in CeGaGe},    
    pdfauthor={Arushi et al.},     
    pdfsubject={},   
    pdfcreator={},   
    pdfproducer={}, 
    pdfkeywords={} {} {}, 
    pdfnewwindow=true,      
    colorlinks=true,       
    linkcolor=blue, 
    citecolor=blue,        
    filecolor=magenta,      
    urlcolor=blue           
} 
\usepackage[normalem]{ulem}

\newcommand{\figref}[1]{Fig.~\ref{#1}}

\renewcommand{\approx}{\simeq}

\makeatletter
\def\maketitle{
\@author@finish
\title@column\titleblock@produce
\suppressfloats[t]}
\makeatother

\begin{document}
\title{Chiral Weyl-Kondo semimetallic state through enhanced correlation in CeGaGe}
\author{Arushi}
\affiliation{Department of Physics and Astronomy, Rice University, Houston, TX, 77005 USA}
\affiliation{Rice Center for Quantum Materials, Rice University, Houston, TX, 77005 USA}
\author{Kevin Allen}
\affiliation{Department of Physics and Astronomy, Rice University, Houston, TX, 77005 USA}
\affiliation{Rice Center for Quantum Materials, Rice University, Houston, TX, 77005 USA}
\author{Yuan Fang}
\affiliation{Department of Physics and Astronomy, Rice University, Houston, TX, 77005 USA}
\affiliation{Rice Center for Quantum Materials, Rice University, Houston, TX, 77005 USA}
\affiliation{Extreme Quantum Materials Alliance, Rice University, Houston, TX, 77005 USA}
\author{Kuan-Sen Lin}
\affiliation{Department of Physics and Astronomy, Rice University, Houston, TX, 77005 USA}
\affiliation{Rice Center for Quantum Materials, Rice University, Houston, TX, 77005 USA}
\affiliation{Extreme Quantum Materials Alliance, Rice University, Houston, TX, 77005 USA}
\author{Mounica Mahankali}
\affiliation{Department of Physics and Astronomy, Rice University, Houston, TX, 77005 USA}
\affiliation{Rice Center for Quantum Materials, Rice University, Houston, TX, 77005 USA}
\affiliation{Extreme Quantum Materials Alliance, Rice University, Houston, TX, 77005 USA}
\author{Hari Bhandari}
\affiliation{Department of Physics and Astronomy, Rice University, Houston, TX, 77005 USA}
\affiliation{Rice Center for Quantum Materials, Rice University, Houston, TX, 77005 USA}
\author{Karthik Rao}
\affiliation{Department of Physics and Astronomy, Rice University, Houston, TX, 77005 USA}
\affiliation{Rice Center for Quantum Materials, Rice University, Houston, TX, 77005 USA}
\affiliation{Rice Laboratory for Emergent Magnetic Materials, Rice University, Houston, Texas 77005, USA}
\author{Alberto Ruiz Biestro}
\affiliation{Department of Physics and Astronomy, Rice University, Houston, TX, 77005 USA}
\author{Christopher Lane}
\affiliation{Theoretical Division, Los Alamos National Laboratory, NM, 87545 USA}
\author{Jian-Xin Zhu}
\affiliation{Theoretical Division, Los Alamos National Laboratory, NM, 87545 USA}
\author{Sanu Mishra}
\affiliation{Department of Physics and Astronomy, Rice University, Houston, TX, 77005 USA}
\affiliation{Rice Center for Quantum Materials, Rice University, Houston, TX, 77005 USA}
\author{Geoffroy Hautier}
\affiliation{Department of Physics and Astronomy, Rice University, Houston, TX, 77005 USA}
\affiliation{Rice Center for Quantum Materials, Rice University, Houston, TX, 77005 USA}
\affiliation{Department of Materials Science and Nanoengineering and Rice Advanced Materials Institute, Rice University, Houston, TX, 77005 USA}
\affiliation{Thayer School of Engineering, Dartmouth College, Hanover NH, 03755 USA}
\author{Qimiao Si}
\affiliation{Department of Physics and Astronomy, Rice University, Houston, TX, 77005 USA}
\affiliation{Rice Center for Quantum Materials, Rice University, Houston, TX, 77005 USA}
\affiliation{Extreme Quantum Materials Alliance, Rice University, Houston, TX, 77005 USA}
\affiliation{Rice Laboratory for Emergent Magnetic Materials, Rice University, Houston, Texas 77005, USA}
\author{Emilia Morosan}
\email[]{emorosan@rice.edu}
\affiliation{Department of Physics and Astronomy, Rice University, Houston, TX, 77005 USA}
\affiliation{Rice Center for Quantum Materials, Rice University, Houston, TX, 77005 USA}
\affiliation{Rice Laboratory for Emergent Magnetic Materials, Rice University, Houston, Texas 77005, USA}
\date{\today}
\begin{abstract}
\begin{flushleft}
\end{flushleft}

Strongly correlated chiral materials have been proposed to host a chiral Weyl-Kondo semimetal (cWKSM) state, in which Kondo hybridization pins chirality-induced Kramers-Weyl point crossings, together with additional symmetry-enforced band crossings, near the Fermi level. Realizing this state requires a material that combines crystal chirality, non-symmorphic symmetry, and Kondo correlations, and currently there are no known materials that combine all these requirements. Here we report evidence for a cWKSM state in CeGaGe, a member of the \textit{R}XY (\textit{R} = rare earth; X = Al, Si; Y = Ga, Ge) family. Magnetization measurements confirm long-range antiferromagnetic order below T$_N$ = 4.7 K with a complex, canted magnetic structure. Unique to CeGaGe among the known \textit{R}XY systems is a structural transition from an achiral tetragonal I4$_{1}$md structure (at high temperatures) to the chiral tetragonal P4$_{3}$ structure (at low temperatures). The structural transition in CeGaGe allows us to highlight the role of chirality in stabilizing the Kramers cWKSM state. Hall resistivity measurements with $H \parallel c$ reveal an anomalous Hall conductivity (AHC) that is constant below the Kondo temperature $T_K \approx 7$ K and drops sharply above it. This crossover tracks the onset of Kondo coherence rather than the magnetic ordering, which occurs at T$_N$ = 4.7 K. Together with first-principles calculations for the $P4_3$ structure, this identifies the AHC as an intrinsic, Berry-curvature-driven contribution generated by Kondo hybridization, rather than a consequence of extrinsic scattering or dynamic scalar spin chirality. These results establish CeGaGe as a rare experimental platform in which crystal chirality, strong electronic correlations, and non-trivial band topology coexist, providing the first material realization of a Kondo-driven cWKSM state.

\end{abstract}
\maketitle

\section*{Introduction}
Topology and strong correlations, once explored largely in isolation, have recently seen a convergence into one of the most active frontiers in condensed matter physics, catalyzed by the discovery of Weyl-Kondo semimetal physics in topological Kondo lattice models theoretically and in the non-magnetic material Ce$_{3}$Bi$_{4}$Pd$_{3}$ \cite{Ce3Bi4Pd3_non-symmorphic_1, Ce3Bi4Pd3_non-symmorphic_2, Ce3Bi4Pd3_non-symmorphic_3, Ce2Au3In5_crystalline_symmetry, Ce3Bi4Pd3_Giant_AHE_exp}. The role of specific crystal symmetries as an organizing principle behind such non-trivial topological features, including Weyl points, nodal lines, and Dirac points, is by now well-established, with non-symmorphic symmetry in both centrosymmetric and non-centrosymmetric structures offering a particularly effective route to non-trivial band crossings \cite{Ce3Bi4Pd3_non-symmorphic_1, Ce3Bi4Pd3_non-symmorphic_3, CeCoGe3_theory_nodal line_and_movingclosetoEf, CeNiSn_non-symmorphic, CeNiSn_CeRhAs_CeRhSb_non-symmorphic, Ce3Pt3Bi4_Ce3Pd3Bi4, CeRu4Sn6}. This symmetry-based picture, however, applies equally well to weakly correlated systems. What makes strongly correlated systems distinctly compelling is the additional role played by the Kondo effect, which acts as an equally critical tuning parameter, pinning these symmetry-enforced crossings close to the Fermi level \cite{Ce3Bi4Pd3_non-symmorphic_2, CeCoGe3_theory_nodal line_and_movingclosetoEf} and thereby bringing them within reach of experimental probes such as magnetotransport of charge and heat, \cite{Ce3Bi4Pd3_Giant_AHE_exp, Magneto_transport_CeCo1-xFexGe3_1, Magneto_transport_CeCo2As2, Magneto_transport_CeCrGe3, Magneto_transport_CeCoGe3_1_and_low gamma_2, UCo0.8Ru0.2Al_Largest_nerst_magneto_tarnsport} and angle-resolved photoemission spectroscopy (ARPES) \cite{CeCoGe3_ARPES, USbTe_ARPES, La1-xCexCo2As2_ARPES, CeNiSn_APRES, CeRhSb_APRES}.

More recently, particular attention has turned to how crystal chirality, defined by the absence of orientation-reversing symmetries, governs non-trivial band crossings in Kondo-correlated materials \cite{cKWKSM, cWKSM_78}. The resulting chiral Weyl-Kondo semimetal (cWKSM) features Kondo-driven Kramers-Weyl nodes at time-reversal invariant momenta (TRIMs), in addition to other topological crossings. These can give rise to novel responses, including the circular photogalvanic effect \cite{photogalvanic_1, photogalvanic_2} and a longitudinal magneto-electric response \cite{magnetoelectric_1, magnetoelectric_2, magnetoelectric_3}. Correlations arising from the Kondo effect pin the 4$f$-derived bands and Weyl points near the Fermi level, in essence supplying a tuning knob that weakly correlated chiral systems lack. However, an experimental realization of this chiral Weyl-Kondo scenario with Kramers-Weyl fermions has been missing to date. The only related report is on Ce$_{3}$Ru$_{4}$Sn$_{13}$ \cite{Ce3Ru4Sn13}, a chiral Kondo lattice compound in which crystal chirality has been mentioned, without a connection to non-trivial band topology.

CeGaGe is established here as the first experimental realization of a Weyl-Kondo semimetal with Kramers-Weyl fermions in a chiral structure, defining it as a Kramers cWKSM. Chirality emerges through a structural transition from the high temperature achiral tetragonal phase to the low temperature chiral tetragonal phase \cite{Structural_transition_CeGaGe}. Related RXY compounds (\textit{R} = rare-earth, X,Y = metal or metalloid) crystallize in the $I4_{1}md$ structure for light \cite{RXY_crystal_structure_LRE1}, and in $Cmcm$ for heavy \textit{R }members \cite{RXY_crystal_structure_HRE1, RXY_crystal_structure_HRE2, RXY_crystal_structure_HRE3, RXY_crystal_structure_HRE4, RXY_crystal_structure_HRE5}, but CeGaGe alone undergoes this structural transition. 


\begin{figure*}[t!]
\includegraphics[width=2.0\columnwidth]{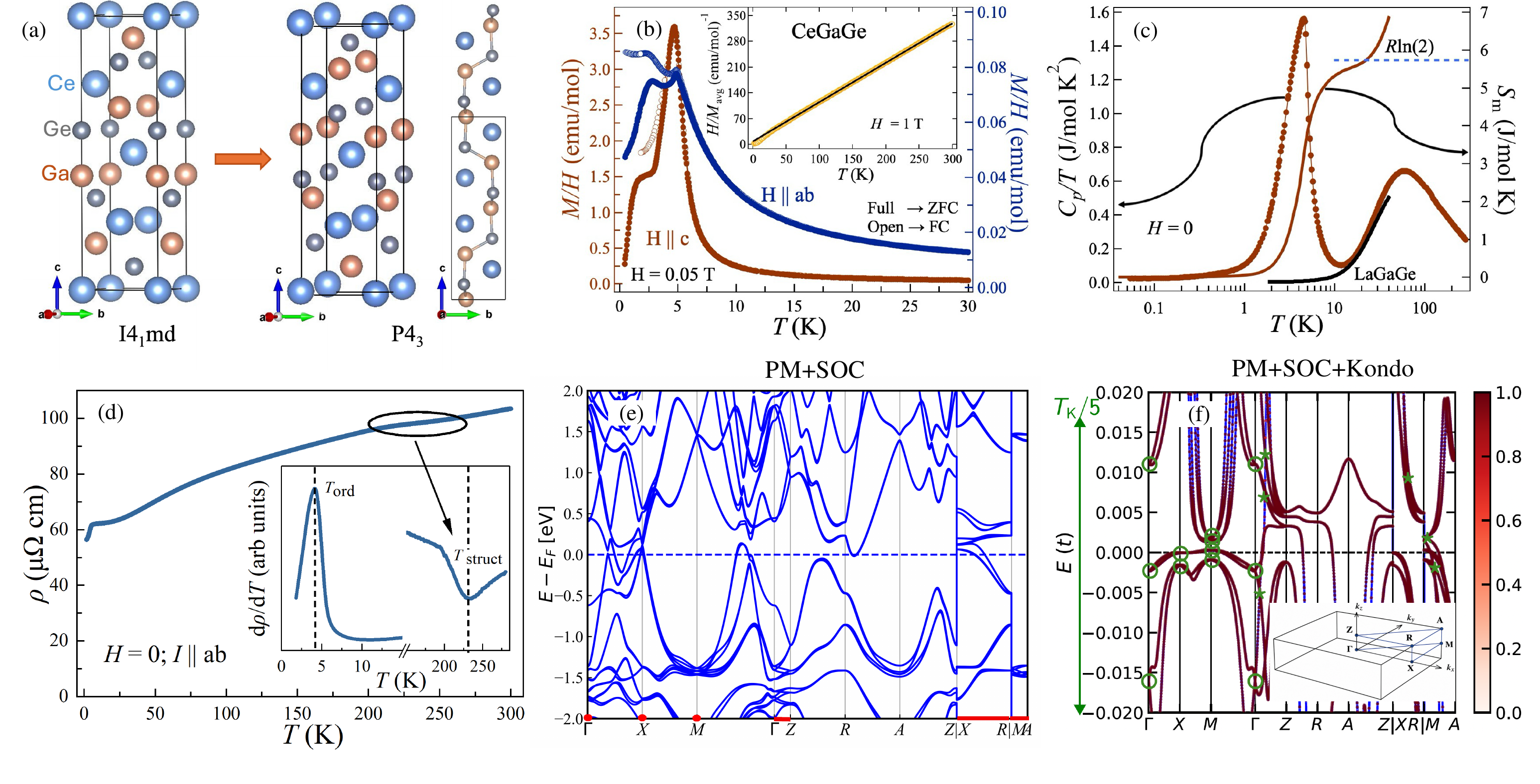} 
\caption{\label{Fig1_new} \textbf{Crystal structure, magnetic ordering, Kondo correlations and electronic band structure of CeGaGe} (a) The high temperature I4$_{1}$md and low temperature P4$_{3}$  crystal structure for CeGaGe, with the right-handed chirality along the c axis for the latter shown on the right. (b) Temperature dependence of the magnetic susceptibility M/H in the low-temperature regime for $H \parallel ab$ (blue) and $H \parallel c$ (red) under an applied field H = 0.05 T (zero-field cooled [ZFC] - full symbols, field cooled [FC] - open symbols. Inset: $H/M_{avg}$ (yellow symbols) (with $M_{avg}$ = [2$M_{ab}$+$M_{c}$]/3) for H = 1 T, with the Curie-Weiss fit shown by the black line. (c) Left axis: Semi-log plot of the H = 0 temperature dependence of the specific heat C$_{p}$/T (red full circles for CeGaGe and black solid line for the non-magnetic analog LaGaGe). Right axis: temperature dependence of the magnetic entropy S$_m$. (d) H = 0 temperature dependence of the electrical resistivity $\rho(T)$ with current $I \parallel ab$; inset: d$\rho/dT$ showing the structural transition at $T_{struct}$ = 230 K and the magnetic ordering at $T_{\mathrm{N}}$ = 4.5 K. (e) First-principles DFT band structure calculations for the low-temperature P4$_{3}$ structure without 4$f$ bands, in the paramagnetic (PM) state with spin-orbit coupling (SOC) included. The high symmetry paths expected to host symmetry-enforced crossings are represented by red horizontal lines, while the TRIM points hosting Kramers-Weyl point crossings are marked by red dots. (f) Renormalized bands from the toy model reproducing the DFT-derived band features, with 4$f$ states capturing the Kondo correlations (strength shown by the color bar) and with SOC included. Energy units are approximate, with t $\approx$ eV. Circles and crosses represent the Kramers-Weyl points and other symmetry-protected crossings near the Fermi level. Inset: the Brillouin zone with the high symmetry path.}
\end{figure*} 

Within this low temperature structure, we report the realization of a Kramers cWKSM state near the Fermi energy $E_{\mathrm{F}}$ in CeGaGe, established through detailed magneto-transport measurements combined with theoretical studies. Theoretically, the DFT-determined band structure was used as a guide to construct a Kondo lattice model, which was then employed to analyze the consequences of the chiral space group no. 78 ($P4_3$) on the Kondo-driven Kramers and other Weyl nodal features. The long-range magnetic order with $T_{\mathrm{N}}$ = 4.9 K is confirmed by magnetization data along both $H$ $\parallel$ $ab$ and $H$ $\parallel$ $c$. The ground state consists of an antiferromagnetic structure, with a weak net ferromagnetic component attributed to the Dzyaloshinskii-Moriya interactions (DMI), on top of the incommensurate order, determined by prior powder neutron diffraction measurements \cite{powder-neutron-diffraction-on-CeGaGe}. Despite the magnetically-ordered ground state, the spin-polarized paramagnetic state along $H$ $\parallel$ $c$ is achieved at a very small field H $\approx$ 0.3 T. This makes CeGaGe a favorable candidate for probing the anomalous Hall conductivity (AHC) in the Zeeman coupled paramagnetic regime, a regime that was theoretically advanced in Kondo lattice models \cite{cKWKSM}. The AHC $\sigma_{xy}^{A}$ with I $\parallel$ $ab$ and $H$ $\parallel$ $c$ and for temperatures below 7 K (close to $T_{\mathrm{K}}$) is constant (independent of $\sigma_{xx}$) $\sigma_{xy}^{A}~=$ 500 $\Omega^{-1}$ $cm^{-1}$, consistent with intrinsic behavior, as is also suggested by theoretical AHC calculations. This is indeed the behavior expected for a cWKSM state. A spontaneous Hall effect is also observed for the same current and field orientation, linked to the magnetically ordered state but decoupled from the net magnetization, and the potential mechanisms behind this behavior are discussed below.


\section*{Results}
\noindent\textbf{Mechanism for the Kramers cWKSM state formation}
\vspace{0.5em}

CeGaGe has been reported to undergo a structural transition between 100 K and 300 K, from the high temperature I4$_{1}$md structure to the lower temperature P4$_{3}$ structure \cite{powder-neutron-diffraction-on-CeGaGe}. However, the exact transition temperature had not been determined. More importantly, in the previous report \cite{CeGaGe_AHE}, band structure calculations were performed for the room temperature I4$_{1}$md structure, while the magnetotransport measurements were carried out at low temperature, where P4$_{3}$ is the relevant phase instead. This mismatch led to a misinterpretation of the transport results, which should be reconciled through calculations and measurements performed for the same structural phase P4$_{3}$ (low temperature).

The CeGaGe structural transition from the I4$_{1}$md to P4$_{3}$ involves only a subtle re-arrangement of atoms (\figref{Fig1_new}a), with the overall tetragonal symmetry preserved. The comparison of the two structures using a translation of the P4$_{3}$ lattice (middle panel of \figref{Fig1_new}a) reveals a slight change in the Ga and Ge atomic positions. The chirality for the P4$_{3}$ structure is apparent from the right-handed Ga-Ge helix along the $c$ axis, shown in \figref{Fig1_new}a (right panel).

Magnetic ordering within this chiral structure of CeGaGe is confirmed by magnetic susceptibility measurements (Fig. \ref{Fig1_new}b). A Curie-Weiss fit of the $H$ = 1 T inverse average susceptibility (\figref{Fig1_new}b, inset) yields an effective moment $\mu_{eff}$ = 2.7 $\mu_{B}$, consistent with the theoretical value for Ce$^{3+}$ with $\mu^{th}_{eff}$ = 2.54 $\mu_{B}$. Zero-field-cooled (ZFC) and field-cooled (FC) $H$ = 0.05 T susceptibility reveals a well defined cusp at $T_{\mathrm{N}}$ = 4.9 K for both for $H$ $\parallel$ $ab$ (blue) and $H$ $\parallel$ $c$ (red). The cusp at $T_{\mathrm{N}}$ suggests AFM correlations, but a ZFC - FC splitting at lower temperatures indicates an additional weak ferromagnetic component. AC susceptibility (Fig. S2b) shows no frequency dependence, ruling out spin freezing or glassy behavior. 

The magnetic order is confirmed by specific heat and resistivity measurements, which also provide evidence for Kondo correlations at low temperatures. The $H$ = 0 specific heat $C_p$(T) for CeGaGe (\figref{Fig1_new}c, full circles) shows a broad Schottky anomaly around 50 K due to the first crystal field excitation. At lower temperatures, a lambda type anomaly at $T_{\mathrm{N}}$ = 4.9 K confirms the long range magnetic order, consistent with magnetization data. A reliable extraction of the electronic specific heat coefficient $\gamma$ is difficult for CeGaGe because the Schottky contribution of the crystal electric field (CEF) extends into the fitting window above $T_{\mathrm{N}}$. Moreover, $C_p$/T \textit{vs.} T develops an anomalous linear dependence below 1 K, whose microscopic origin remains undetermined and cannot be captured using standard electronic or magnonic terms. Bounding estimates of $\gamma$ are obtained from the low temperature plateau (below 0.2 K) and from the high temperature fit (11 - 20 K), yielding $\gamma$ $\approx$ 27 mJ/mol-K$^{2}$, and $\approx$ 45 mJ/mol-K$^{2}$, respectively. Both values exceed that of the non-4$f$ analog LaGaGe $\gamma$(LaGaGe) = 1.7 mJ/mol-K$^{2}$, suggesting enhanced electronic correlations in CeGaGe. This places CeGaGe among other moderately correlated Ce-based compounds such as CeNiGe$_{3}$ \cite{CeNiGe3_low gamma_1}, nodal-line semimetal CeCoGe$_{3}$ \cite{CeCoGe3_low gamma_1, Magneto_transport_CeCoGe3_1_and_low gamma_2}, CePtSi$_{3}$ \cite{CePtSi3_low gamma}, CeRhGe$_{3}$ \cite{CeRhGe3_low gamma, CeRhGe3_low gamma_1}, CeAgAs$_{2}$ \cite{CeAgAs2_low gamma}, CeCuSi \cite{CeCuSi_low gamma}, and a well-studied correlated topological compound CeNiSn \cite{CeNiSn_low gamma_1, CeNiSn_low gamma_2}. A $\gamma$ = 50 mJ/mol-K$^{2}$ is also reported for CeAlGe \cite{spinflop_CeAlGe_plus_schootky_anomaly}, however, the implications of correlations in this compound was not discussed. The magnetic entropy S$_{m}$(T) for CeGaGe (\figref{Fig1_new}c) obtained after subtracting the phonon contribution using LaGaGe, shows that the full doublet entropy Rln(2) is recovered at $T$ = 23 K, with only 60\% entropy at $T_{\mathrm{N}}$, consistent with Kondo screening. Using S$_m(T_{K}/2)~=~\frac{1}{2}R ln2$, the Kondo temperature is estimated as $T_K~\approx$ 8 K, close to $T_{\mathrm{N}}$. Together, the enhanced $\gamma$ and reduced entropy at $T_{\mathrm{N}}$ provide evidence for Kondo correlations in CeGaGe but with relatively weak $c-f$ hybridization. Further details about the specific heat fitting of LaGaGe, and $\gamma$ extraction of CeGaGe are provided in the Supplementary Material Sec. 4.
 
The structural transition is best observed in the temperature dependent resistivity $\rho$(T), as a change in slope circled in \figref{Fig1_new}d, and is resolved more clearly as a kink at T$_{struct}$ = 230 K in $d\rho$/$d$T (inset, right). This is indeed consistent with the neutron diffraction estimate, which places the transition between 100 K and 300 K  \cite{Structural_transition_CeGaGe}. Due to the loss of spin-disorder scattering expected at the magnetic ordering temperature, T$_N$ is further confirmed by a drop in $\rho(T)$ at low temperatures (\figref{Fig1_new}d). This results in a sharp peak in $d\rho/dT$ (inset, left) at $T_{\mathrm{N}}$ = 4.5 K, consistent with magnetization and specific heat measurements. $\rho$(T) also displays a broad hump from 50 K to 200 K, attributed to Kondo scattering of excited CEF levels, as reported for other Cerium-based Kondo lattice systems \cite{CeNiGe3_low gamma_1}. In Cerium-based compounds with magnetic order, a clear logarithmic Kondo upturn in $\rho$(T) is often absent when $T_{\mathrm{N}}$ and $T_{\mathrm{K}}$ are comparable, since the onset of magnetic order tempers the Kondo signature. This is indeed the case in CeGaGe, where $T_{\mathrm{N}}$ = 4.9 K and T$_{K}~\approx$ 8 K are close. However, Kondo correlations are evident from the field dependence of $\rho$(T) for $H$ $\parallel$ $c$ (Fig. S3a of Supplementary Material Sec. 3). 

A field of $H$ = 0.3 T ($H$ $\parallel$ $c$) suppresses the magnetic order below $T$ = 1.8 K. When applying a slightly larger field ($H$ = 0.5 T), the broad hump at low temperatures persists, then weakens with increasing field and vanishes above $H$ = 11 T (Supplementary Material Fig. S3a). This behavior reaffirms the presence of Kondo correlations in CeGaGe, which was first indicated by the low entropy (60\%) at $T_{\mathrm{N}}$. 

The structural transition in CeGaGe provides a unique route to examine how a chiral crystal structure impacts non-trivial topology when combined with Kondo correlations. Such a transition in CeGaGe is distinct and clean, offering an advantage over the alternative scenario of two distinct compounds, each crystallizing in one of the two structures of CeGaGe. Disorder and crystal quality can vary between samples and influence carrier density, while SOC, chemical potential, and orbital character near the Fermi level can also differ, complicating any direct comparison. The structural transition within a single compound, as in CeGaGe, therefore offers a cleaner route, isolating the space group change as the sole variable responsible for the observed behavior.

\begin{figure*}[t!]
\includegraphics[width=2.0\columnwidth]{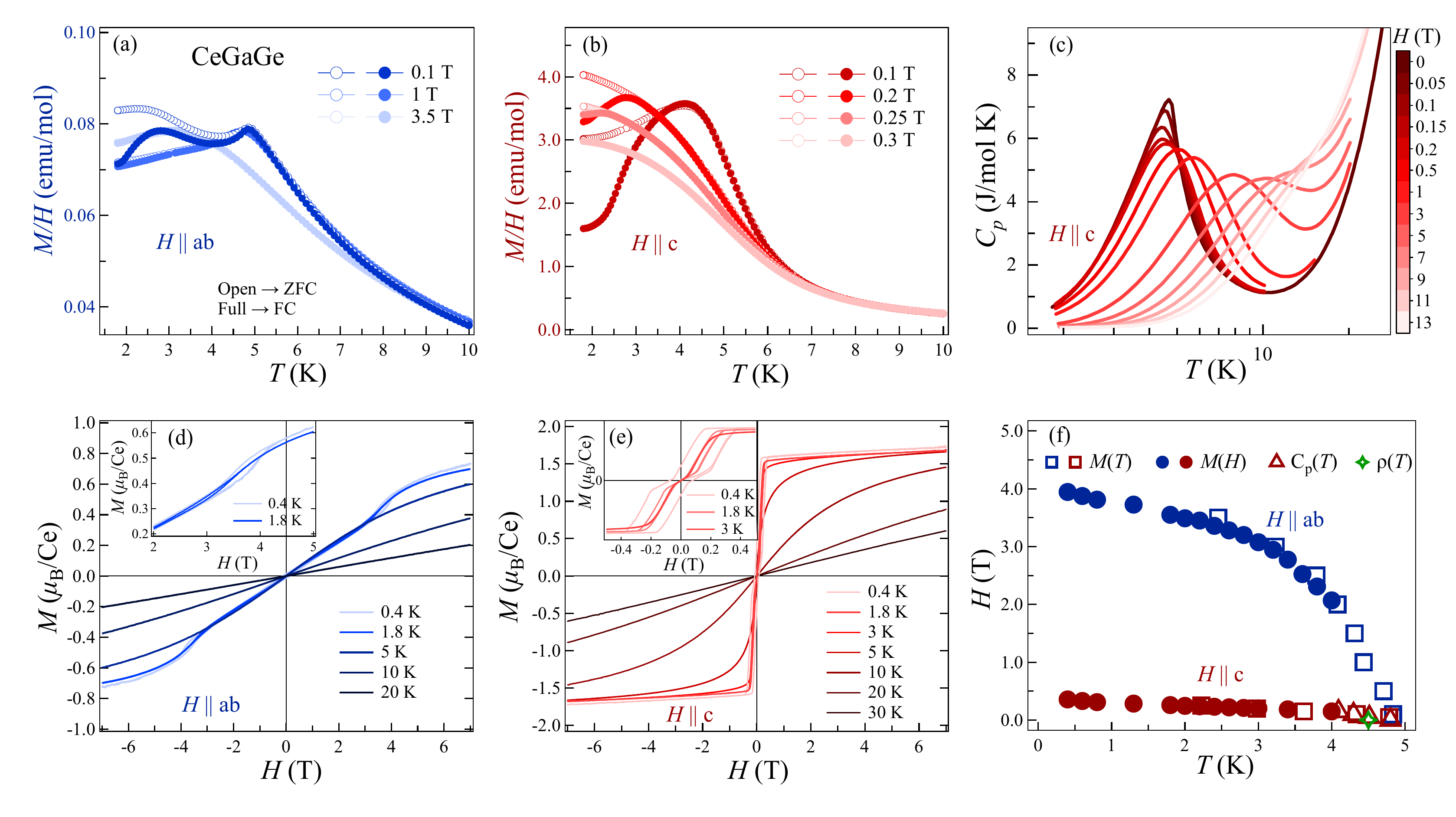} 
\caption{\label{Fig2_new} \textbf{H - T phase diagram derived from thermodynamic and transport measurements} (a, b) Magnetic susceptibility for $H$ $\parallel$ $ab$ (blue) and $H$ $\parallel$ $c$ (red) for H = 0.1 T - 3.5 T. (c) Specific heat C$_{p}$ for $H$ $\parallel$ $c$. (d) Magnetization isotherms for $H$ $\parallel$ $ab$. Inset: a zoomed in view of the hysteresis at the metamagnetic transition for T = 0.4 K - 1.8 K. (e) Magnetization isotherms for $H$ $\parallel$ $c$. Inset: a zoomed in low-field view of the evolution of M($H$) from 0.4 K to 3 K. (f) H - T phase diagram for $H$ $\parallel$ $ab$ (blue) and $H$ $\parallel$ $c$ (red), constructed from M($T$), C$_{p}$($T$), $\rho(T)$ and M($H$) data.}
\end{figure*}


The low temperature space group P4$_{3}$ contains the nonsymmorphic screw rotations \{C$_{4z}$|0,0,3/4\} and \{C$_{2z}$|0,0,1/2\}, corresponding to $\pi$/2 and $\pi$ rotations about the cartesian z axis, respectively. Together with the lack of inversion symmetry and in the presence of SOC, these operations enforce multiple symmetry-protected band crossings in the PM state \cite{symmetry_enforced}: i) Weyl points along the $\Gamma$-Z and M-A high symmetry lines, arising from accordion-type band connectivity enforced by \{C$_{4z}$|0,0,3/4\}; ii) Weyl points along X-R, arising from hourglass-type band connectivity enforced by {C$_{2z}$|0,0,1/2\}; and iii) a Weyl nodal plane at the k$_{z}$ = $\pi$ Brillouin zone boundary, enforced by the composite symmetry $\{C_{2z}\mathcal{T}|0,0,1/2 \}$ where $\mathcal T$ is time-reversal symmetry (TRS) (\figref{Fig1_new}e) (Supplementary Material Sec. 6). The DFT calculated band structure of CeGaGe in the PM state with SOC and without 4$f$ bands (\figref{Fig1_new}e) shows these crossings within the highlighted regions (red lines) along $\Gamma$-Z, M-A, and X-R. Because the room-temperature structure is also reported to host Weyl crossings \cite{I41md_non-trivial_topology}, the main consequence of this structural transition is the introduction of chirality, which allows for additional Weyl crossings, hosting Kramers-Weyl fermions at the TRIM points X, M, and $\Gamma$, inherent to chiral structures \cite{topology_chiral_crystal_Weakly,topology_chiral_crystal_(TaSe4)2I}.

Since Kondo correlations are established, the Kondo driven 4$f$ states must be included in the model. We first construct a toy model from two elementary band representations that captures the essential features of the DFT-derived band features, including the orbital character and
band topology of the states near E$_{\rm F}$ (Supplementary Material Sec. 8). Next, the Kondo effect is taken into account (see the Supplementary Material Sec. 8 and ref. \cite{cKWKSM}, \cite{cWKSM_78}), which leads to 4$f$ bands that are shown (Fig. \ref{Fig1_new}f, red) with weight concentrated near the Fermi level. The resulting Kondo-derived heavy fermion bands are pinned close to E$_{\rm F}$, generating both the Kramers-Weyl points and the symmetry-enforced Weyl points (circles and crosses, respectively) in the immediate vicinity of the Fermi level. This proximity dictates the low temperature physics of CeGaGe, as these Weyl crossings act as sources of Berry curvature. Their contribution can be probed through transport measurements such as AHC, discussed in the magneto-transport section below, and identify CeGaGe as a Kramers cWKSM candidate system.

\vspace{2.0em}
\noindent\textbf{Antiferromagnetic correlations in CeGaGe}
\vspace{0.5em}

Before the transport signatures are addressed, the magnetic characteristics of CeGaGe need to be firmly established, since these underscore both their interpretation and the choice of experimental transport probes.
Neutron powder diffraction experiments \cite{powder-neutron-diffraction-on-CeGaGe} established the magnetic order near $T$ = 5 K as incommensurate. Our temperature-dependent magnetization $M/H$(T) and specific heat measurements $C_p$(T) for different applied fields (\figref{Fig2_new}a-c) show the suppression of $T_{\mathrm{N}}$ with increasing field, as expected for antiferromagnetic order. For $H$ $\parallel$ $ab$, a second transition is apparent in the susceptibility at $T$ = 3.5 K (Fig. \ref{Fig2_new}a). For $H$ $\parallel$ $c$, only the transition at $T_{\mathrm{N}}$ is observed in both $M/H$ and $C_p$. Furthermore, the H $\parallel$ \textit{c} susceptibility is an order of magnitude larger than in the H $\parallel$ \textit{ab} direction, suggesting an easy-axis anisotropy.

The field-dependent magnetization measurements provide further insight into the magnetic order. For $H$ $\parallel$ $ab$ (\figref{Fig2_new}d), M(H) is linear at low fields. At $T$ = 0.4 K, a small hysteresis loop opens up between $H$ = 3.3 T and 4.4 T (inset, \figref{Fig2_new}d) at a metamagnetic transition, consistent with a spin-flop transition \cite{spinflop_CeAlGe_plus_schootky_anomaly, spinflop_Cu0.95MnAs, spinflop_spincanting_Cr2O3}. On warming, the metamagnetic transition persists up to $T$ = 1.8 K but is weaker and shifted to lower fields, and is fully suppressed by $T$ = 5 K. At $H$ = 7 T, M(H) values remain well below the theoretical saturation value of 2.14 $\mu_{B}$ for Ce$^{3+}$. M(H) along the easy axis $H$ $\parallel$ $c$ (\figref{Fig2_new}e) is markedly different. At $T$ = 0.4 K, M increases sharply below $H$ = 0.4 T, and saturates close to 1.71 $\mu_{B}$ at higher fields. The reduced moment relative to the free ion value, together with the unsaturated $H$ $\parallel$ $ab$ response, reflects the combined effects of CEF and Kondo screening in CeGaGe. A closer look at the low field behavior (inset, \figref{Fig2_new}e) reveals a metamagnetic transition around $H$ = 0.2 T ($T$ = 0.4 K), which moves down in field as T increases up to 3 K, and disappears at higher temperatures. Remarkably, the $T$ = 0.4 K M(H) isotherms show hysteresis around $H$ = 0, but the M(H) loops close up at zero field for higher temperatures, even as the metamagnetic transition persists.

Together, the $H$ $\parallel$ $ab$ and $H$ $\parallel$ $c$ data support a magnetic structure in which moments lie predominantly along the $c$ axis, with a small canted in-plane component. The weak ferromagnetic component, second transition in the hard magnetization direction, and the change in M(H) hysteresis behavior at $T$ = 1.8 K for $H$ $\parallel$ $c$, closely parallels the behavior reported in DyScSi \cite{DyScSi}, where powder neutron diffraction identified the higher temperature transition as incommensurate AFM order and the lower one as the onset of a weak ferromagnetic component. The symmetry-allowed DMI for the non-centrosymmetric P4$_{3}$ structure suggests a non-collinear moment arrangement in CeGaGe \cite{PrAlGe}. Complex magnetism is a recurring feature in the isostructural compounds, such as non-collinear order in CeAlSi \cite{CeAlSi}, non-coplanar spin textures in CeAlGe \cite{CeAlGe}, stripe helical magnetism in NdAlGe \cite{NdAlGe}, helical magnetism in NdAlSi \cite{NdAlSi}, spiral magnetism in Kramers Nodal line and Weyl fermion compound SmAlSi \cite{SmAlSi_1, SmAlSi_2}. 

A magnetic phase diagram (\figref{Fig2_new}f) is constructed from the M(T;H), $C_p$(T) and $\rho$(T) data, showing that a much weaker field suppresses the transition for the easy axis (H $\parallel$ $c$, red symbols) compared to the H $\parallel$ $ab$ (blue symbols). Furthermore, the field anisotropy points to a strong easy-axis magnetic anisotropy along $c$, where a very small field along this direction is sufficient to access the spin-polarized PM state. Because this spin-polarized state is achieved at such low fields, it provides access to the AHC in the paramagnetic state, while still offering a sufficient field range to analyze and interpret the data. This establishes $H$ $\parallel$ $c$ as the primary direction for probing the AHC.

\vspace{2.0em}
\noindent\textbf{Anomalous Transport Properties}
\vspace{0.5em} 

To gain more insight into the electronic transport properties, the CeGaGe magnetoresistance 
\begin{equation*}
 \text{MR} = \frac{\rho_{xx}(H)-\rho_{xx}(0)}{\rho_{xx}(0)}
\end{equation*}
was measured for I $\parallel$ $a$ and $H$ $\parallel$ $c$ (the easy axis) for temperatures between 1.8 K and 30 K. 
At $T$ = 1.8 K, MR drops near $H$ $\approx$ 0.25 T (\figref{Fig3_new}b), consistent with the transition observed in the isothermal magnetization M(H). With increasing field, MR reaches a minimum near $H$ = 2 T, followed by a nearly linear increase for higher fields (\figref{Fig3_new}a). The low field negative MR is attributed to the suppression of the spin-flip scattering off magnons, while the high field upturn indicates that the Lorentz-force scattering dominates. A small hysteresis is observed in the ordered state (T $\leq$ 3 K) at low fields, and is absent above $T_{\mathrm{N}}$ (\figref{Fig3_new}b). Although the ground state magnetic structure of CeGaGe is reported as incommensurate, powder neutron diffraction could not resolve the specific structure type or nature of the modulation; nevertheless, the structure is complex enough to support domain walls \cite{TmB4_hystereticMR_Domainwall, Sr2IrO4_hystereticMR_Domainwall}. Therefore, the hysteretic MR is consistent with enhanced spin disorder scattering at these domain walls. Whether the domain walls arise within a single incommensurate ground state, as reported for EuIn$_2$As$_2$ \cite {EuIn2As2_hystereticMR_Domainwall}, or at the boundary between distinct magnetic phases (\textit{e.g.} collinear to helical) as in EuCuSb \cite{EuCuSb_hystereticMR_Domainwall} and EuZnGe \cite{EuZnGe_hystereticMR_Domainwall}, cannot be resolved without single crystal neutron diffraction. The  MR remains negative up to 26 K, well above $T_{\mathrm{N}}$, which can be attributed to the combined effects of Kondo and spin fluctuations scattering.

\begin{figure*}[t!]
\includegraphics[width=2.0\columnwidth]{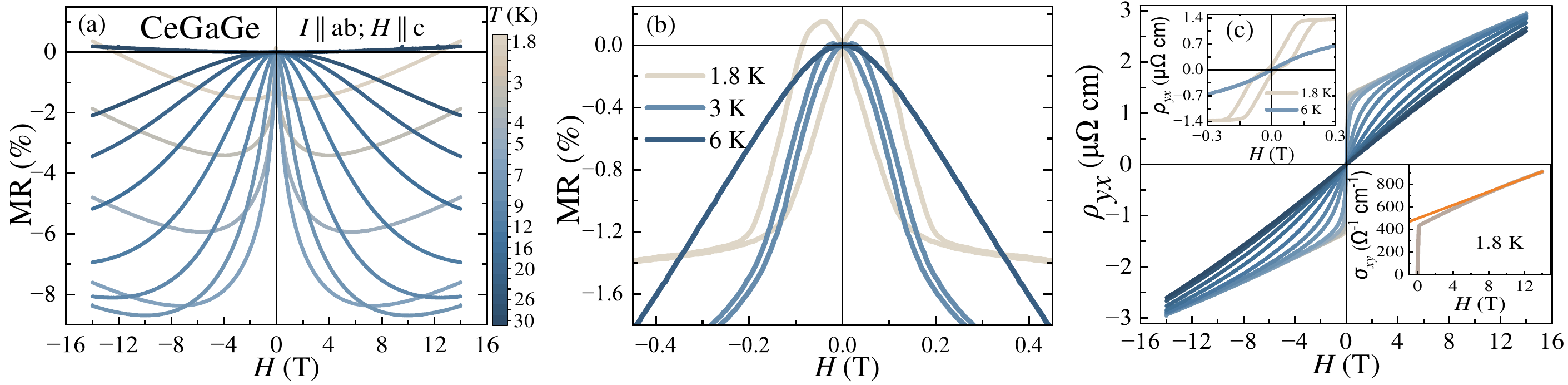} 
\caption{\label{Fig3_new} \textbf{$H \parallel c$ magnetotransport measurements} (a) Magnetoresistance MR for current $I \parallel ab$ and $H$ $\parallel$ $c$, for temperatures from 1.8 K to 30 K. (b) Selected MR isotherms illustrate the evolution of hysteresis at low fields: maximum at T = 1.8 K, diminishes at 3 K, and disappears entirely at 6 K (above $T_{\mathrm{N}}$). (c) Hall resistivity $\rho_{yx}$ at the same temperatures as in the MR measurements. The upper left inset shows a zoomed-in view of the low field regime, revealing the spontaneous Hall effect at T = 1.8 K and its absence at 6 K (above $T_{\mathrm{N}}$). The lower right inset displays the Hall conductivity at T = 1.8 K, with a linear fit (orange line) for the high-field regime, extrapolated to H = 0 to extract the anomalous Hall conductivity.}
\end{figure*}

When we turn to Hall resistivity measurements $\rho_{yx}$ performed using the same current and field configuration as the MR measurements (\figref{Fig3_new}c), we find that $\rho_{yx}$ deviates from linearity both below and above $T_{\mathrm{N}}$. This field dependence can be attributed to multi-band effects and non-trivial band topology. However, the linear $\rho_{yx}$ at $T$ = 100 K (shown in Fig. S3b of Supplementary Material Sec. 3) rules out a multi-band origin, which leaves non-trivial topology as the dominant contribution. The total Hall resistivity for CeGaGe can be expressed as a sum of ordinary Hall effect (OHE) and anomalous Hall effect (AHE) contributions:
\begin{equation*}
 \rho_{yx}^{tot} = \rho_{yx}^{ord}+\rho_{yx}^{A}
\end{equation*}
with $\rho_{yx}^{ord}$ = R$_{0}$$\mu_{0}$H, and $\rho_{yx}^{A}\propto$ M can originate from either intrinsic or extrinsic mechanisms. The intrinsic mechanism was first proposed by Karplus and Luttinger, who attributed it to the SOC \cite{KL_intrinsic_use_ref for converting AHE to AHC_3}. This has since been reformulated as a scattering-independent Berry curvature of occupied Bloch states, acting as a fictitious momentum space magnetic field. The extrinsic mechanisms, skew scattering and side jump scattering, both arise from SOC-mediated electron scattering, with the former dominating in a cleaner, high longitudinal conductivity regime, and the latter most relevant in the bad-metal regime. To isolate the anomalous contribution, $\rho_{yx}$ was first converted to Hall conductivity via $\sigma_{xy}$ = $\rho_{yx}$/$\rho_{xx}^{2}$ (valid for $\rho_{yx}$ $<<$ $\rho_{xx}$) \cite{ref for converting AHE to AHC_1, ref for converting AHE to AHC_2, ref for converting AHE to AHC_3, high conductivity leading to skew scattering regime_use ref for converting AHE to AHC as well}, shown for $T$ = 1.8 K in the inset of \figref{Fig3_new}c. A linear fit in the spin-polarized state (9 T < H < 14 T) was extrapolated to zero field, following the approach used for other topologically non-trivial magnetic compounds \cite{other_nontrivial_topological_systems}. This yields $\sigma_{xy}^{A}$ $\approx$ 500 $\ohm^{-1}$cm$^{-1}$ at $T$ = 1.8 K, placing it firmly within the intrinsic regime of AHC. Repeating this procedure up to $T$ = 30 K provides $\sigma_{xy}^{A}(T)$, which are plotted as a function of the longitudinal conductivity $\sigma_{xx}$, shown in \figref{Fig4_new}a (red hexagons). Interestingly, below $T$ = 7 K, $\sigma_{xy}^{A}$ is independent of $\sigma_{xx}$, a hallmark of the intrinsic Berry curvature mechanism. Furthermore, the $\sigma_{xx}$ values lie in the intermediate good-metal regime (10$^{4}$ - 10$^{6}$ $\ohm^{-1}$cm$^{-1}$), well below the ultra clean limit (> 10$^{6}$ $\ohm^{-1}$cm$^{-1}$) where skew scattering would be expected to dominate, providing further evidence that the intrinsic mechanism is the primary contributor to the AHE. However, because $\sigma_{xy}^{A}$ is independent of $\sigma_{xx}$ below T$_K$ and rapidly decreases above T$_K$ suggests that extrinsic contributions may exist at higher temperatures. The anomalous Hall angle (AHA) is calculated at the lowest temperature, which is 1.61$\degree$, consistent with most magnetic compounds where AHA ($\theta_{A}$) < 3$\degree$ \cite{AHA} and Weyl-semimetal candidate CeAlSi \cite{AHA_CeAlSi}. For temperatures between 7 K and 20 K, $\sigma_{xy}^{A}$ remains larger than the extrinsic contributions, and its rapid increase with $\sigma_{xx}$ on cooling in this temperature range is also inconsistent with conventional skew scattering from static, spin-orbit coupled impurities. Instead, this points to thermally evolving scattering sources, such as spin fluctuations above $T_{\mathrm{N}}$ or the onset of Kondo coherence. Above 20 K, $\sigma_{xy}^{A}$ approaches saturation, consistent with a conventional extrinsic mechanism dominating in this regime. A comparison of the anomalous Hall conductivity for CeGaGe with other correlated topologically non-trivial (\figref{Fig4_new}b) reinforces the origin of large Berry curvature from non-trivial band crossings near E$_{\rm F}$.   
 
 
\begin{figure*}
\includegraphics[width=2.0\columnwidth]{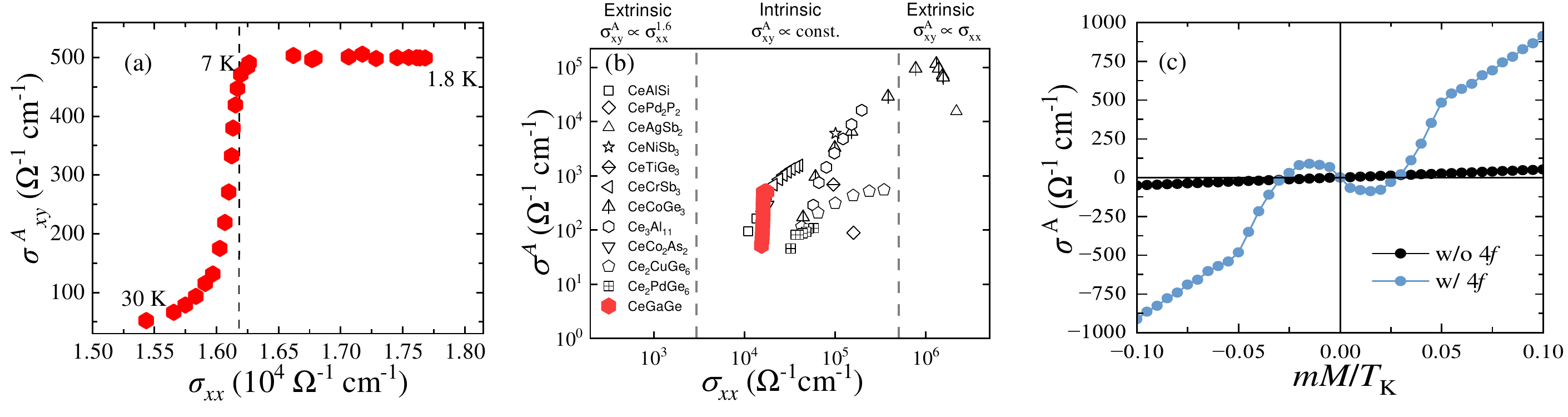} 
\caption{\label{Fig4_new} \textbf{Anomalous Hall conductivity $\sigma^A_{xy}$ as a function of the longitudinal conductivity and theory comparison.} (a) $\sigma^{A}_{xy}$ as a function of $\sigma_{xx}$ from 1.8 K to 30 K. Below 7 K (vertical line) $\sigma^{A}_{xy}$ remains constant with respect to $\sigma_{xx}$, consistent with an intrinsic contribution. (b) Comparison of $\sigma^{A}$ $vs$ $\sigma_{xx}$ for CeGaGe (red symbols) and other topologically non-trivial Ce-based compounds (open symbols, black). \cite{scaling-plot_1, scaling-plot_3, Magneto_transport_CeCoGe3_1_and_low gamma_2, scaling-plot_5, scaling-plot_6, Ce3Al11, Ce2Cu_PdGe6}. (c) Theoretical $\sigma^{A}$ as a function of effective Zeeman field $mM/T_\mathrm{K}$ for the toy model (Supplementary Material Sec. 8) with 4$f$ bands (blue) and without 4$f$ bands (black), respectively. Large nonzero values are obtained when 4$f$ bands are included, while near zero values result when 4$f$ bands are excluded.}
\end{figure*}

As shown in \figref{Fig1_new}, the chiral, nonsymmorphic space group P4$_{3}$ hosts an intrinsic source of Berry curvature through symmetry-enforced band crossings and Kramers-Weyl point crossings, though in the bare band structure these lie far away from E$_{\rm F}$. The Kondo coupling between Ce 4$f$ and conduction electrons produces resonance excitations with a narrow energy dispersion and pins the resulting heavy quasiparticle bands close to E$_{\rm F}$ (\figref{Fig1_new}f). As a result, both types of crossings are develop near the Fermi energy, a scenario associated with the Weyl-Kondo semimetals \cite{Ce3Bi4Pd3_non-symmorphic_1, Ce3Bi4Pd3_non-symmorphic_2, Ce2Au3In5_crystalline_symmetry}, which are evidenced in heavy fermion semimetals such as Ce$_{3}$Bi$_{4}$Pd$_{3}$ \cite{Ce3Bi4Pd3_non-symmorphic_2, Ce3Bi4Pd3_Giant_AHE_exp} and CeRu$_{4}$Sn$_{6}$ \cite{CeRu4Sn6}. The $\sigma_{xy}^{A}$ independent of $\sigma_{xx}$ below 7 K, close to $T_{\mathrm{K}}$ and above $T_{\mathrm{N}}$, points to Kondo coherence rather than magnetic order as the relevant energy scale organizing this Berry curvature, positioning CeGaGe as a cWKSM candidate. Although dynamic scalar spin chirality has been proposed as a possible source of AHE in the PM state, its contribution in CeGaGe remains experimentally unverified. The crossover from intrinsic to extrinsic AHC occurs near 7 K, close to T$_K$ and definitely higher than T$_N$. Therefore, this crossover is independent of the magnetic order. Moreover, M(H) remains non-linear even at 10 K, well above the AHC crossover, where a contribution from dynamic scalar spin chirality could also be expected. However, while this mechanism can not be excluded, the AHC crossover close to T$_K$ and above T$_N$ favors a Kondo-driven interpretation of the observed AHE.

To test this scenario, the theoretical AHC was calculated from a low energy toy model reproducing the orbital character and band topology near E$_{\rm F}$ obtained from DFT (Supplementary Material Sec. 8). Since the experimental AHC was extracted by extrapolating $\sigma_{xy}$ from the high field, forced-ferromagnetic regime, the required condition of broken TRS is satisfied by the applied field. Unlike systems in which field-induced TRS breaking splits degenerate points into new Weyl nodes, CeGaGe already hosts Weyl nodes arising from chirality and non-symmorphic crystal symmetry. Instead, the field deforms or splits these pre-existing nodes, discussed below. 

The theoretical AHC as a function of the effective Zeeman field $mM/T_\mathrm{K}$, the model proxy for TRS-breaking magnetization, is shown in \figref{Fig4_new}c. A large, non-zero response appears even at small $mM/T_\mathrm{K}$, but only when the 4$f$ bands are included. Removing them suppresses the response, implicating the Kondo-derived bands in the AHE. To further justify the inclusion of 4$f$ bands, AHC was calculated from the wannierized bands of the DFT calculation, with the 4$f$ bands pushed into the core. The maximum absolute value of AHC obtained from these calculations was close to 160 $\ohm^{-1}$cm$^{-1}$, much smaller than the experimental values at low temperatures (Supplementary Material Sec. 5). Therefore, 4$f$ bands must be included to account for the total AHC value. The theoretical AHC (\figref{Fig4_new}c) is calculated by setting the chemical potential at E$_{\rm F}$, confirms that this large response is tied to crossings pinned at E$_{\rm F}$ and would be strongly altered by doping- or pressure-induced shifts of E$_{\rm F}$. This response arises from Zeeman-induced avoided or deformed crossings, where the former includes nodal planes in the k$_{z}$ = $\pi$ plane, and the latter includes the Kramers-Weyl points at TRIMs, symmetry-enforced Weyl points along the high symmetry line and also accidental Weyl points. A natural question is whether the Kramers-Weyl contribution can be isolated from these other crossings in this Kramers cWKSM candidate. Given the small Kondo scale and the small energy window of the relevant 4$f$-derived crossings, the Berry curvature likely arises collectively from all of them, making an unambiguous decomposition difficult. Therefore, the toy model results support an intrinsic, Kondo-renormalized origin for the observed AHC, consistent with the $\sigma^{A}$-$\sigma_{xx}$ scaling observed experimentally below 7 K ($T_{\mathrm{K}}$). Rather than a single class of band crossings, the collective contribution of Kondo-pinned, chirality- and symmetry-protected crossings near E$_{\rm F}$ firmly establishes CeGaGe as a Kramers cWKSM. 

Another striking feature of the CeGaGe Hall effect at $T$ = 1.8 K is the emergence of a spontaneous (zero-field) Hall signal (\figref{Fig3_new}c). The hysteresis in $\rho_{yx}$ vanishes above 5 K, directly linking it to the magnetically ordered state. A comparison with M(H) at $T$ = 1.8 K reveals clear decoupling between the two responses: M(H) remains strictly linear up to $H$ = 0.1 T, with no hysteresis, while hysteretic opening in M(H) develops only between 0.1 T and 0.28 T. In contrast, the Hall response already exhibits spontaneous hysteretic opening at zero field (\figref{Fig3_new}c), which is precisely within the regime where M(H) is linear and the net moment vanishes. A microscopic origin of the hysteretic behavior of the AHC remains to be established. The exact spin configuration and its field-driven evolution remain undetermined in CeGaGe. However, a likely magnetic configuration can be estimated based on the magnetization and powder neutron diffraction data, suggesting an incommensurate, predominantly antiferromagnetic order with a weak net ferromagnetic component. The complexity of such a magnetic structure makes it natural to consider scenarios discussed in the literature for spontaneous zero-field Hall signals in the absence of net magnetization. A non-coplanar spin arrangement with finite scalar spin chirality can generate a real space fictitious magnetic field and a spontaneous Hall conductivity \cite{SHE-non-coplanar}. In non-collinear configurations, spin-orbit assisted k-space Berry curvature from the magnetic structure can also produce a spontaneous zero field Hall response \cite{SHE-non-collinear1, SHE-non-collinear2}. More generally, antiferromagnetic order that breaks the combined time-reversal and translation symmetry can host a spontaneous Hall effect entirely independent of the net magnetization \cite{SHE-Tt-AFM}. Additionally, classical pinning of domain walls at structural defects can produce a remanent and hysteretic Hall signal whenever the magnetically ordered phase itself supports a finite AHC, without requiring any non-trivial topological contribution. All of these scenarios naturally produce a hysteretic Hall response decoupled from bulk magnetization, consistent with our observations. Identifying the exact mechanism requires determination of the magnetic space group and the field dependent evolution of the incommensurate order parameter by performing single crystal neutron diffraction measurements under applied fields, and this is the subject of an ongoing study.

\section*{Discussion}

CeGaGe is a chiral compound likely to host a Kramers cWKSM state. Kramers-Weyl point crossings are enabled by chirality, in addition to other symmetry-enforced Weyl band crossings present from the underlying crystal symmetry. Both are pinned close to $E_{\rm F}$ by Kondo correlations, where the associated Berry curvature is expected to yield an intrinsic contribution to AHC once Kondo hybridization sets in. This expectation is borne out by Hall measurements for $I \parallel ab$ and $H \parallel c$ (\figref{Fig3_new}c, where a constant AHC of 500 $\Omega^{-1}$cm$^{-1}$ emerges below 7 K, tracking $T_{\mathrm{K}}$ rather than the antiferromagnetic ordering temperature $T_N = 4.7$ K. On cooling, the onset of $\sigma^A_{xy}$ precedes the onset of magnetic order by more than 2 K, and it persists, independent of  $\sigma^{xx}$, in the spin-polarized paramagnetic state reached at $H \approx 0.3$ T along $H \parallel c$. Skew scattering and other extrinsic mechanisms are unlikely since $\sigma^A_{xy}$ is independent of $\sigma_{xx}$ across the Kondo coherence crossover T$_K$, inconsistent with the scaling expected for extrinsic skew-scattering contributions. However, this is consistent with an intrinsic, Berry-curvature-derived AHC governed by Kondo correlations. The coincidence of the AHC crossover with $T_K$ and not with $T_N$ is evidence that Kondo hybridization, not magnetic order, generates the Berry curvature source: as coherent Kondo screening develops below $T_K$, the hybridized $f$-$d$ bands sharpen and pin the Kramers-Weyl and symmetry-enforced Weyl nodes at $E_F$, producing the observed AHC. Above $T_K$, loss of coherence pushes these nodes away from $E_F$ and suppresses $\sigma^A_{xy}$ sharply, (\figref{Fig4_new}a). This picture is consistent with the theoretically-proposed cWKSM state and is further supported by AHC calculations from a toy model constructed from DFT-derived bands. Long-range magnetic order is confirmed by magnetization, specific heat, and resistivity measurements, revealing dominant antiferromagnetic correlations with a weak ferromagnetic component. However, because the intrinsic AHC is isolated in the field-polarized regime well above the ordering field, it is decoupled from this magnetic structure, and instead tracks the Kondo energy scale, providing the first experimental realization of the recently proposed cWKSM state \cite{cKWKSM}. CeGaGe also exhibits a spontaneous Hall response in the ordered state; several possible origins reported for other systems are discussed, though a definitive conclusion requires solving the magnetic structure through single-crystal neutron diffraction, and is left for future work. The structural transition in CeGaGe enables Kramers points through chirality in the low temperature structure (space group 78), and Kondo correlations pin them exactly at $E_F$. CeGaGe is therefore the first system in which an intrinsic, Kondo-driven AHC occurs in a Kramers cWKSM state.

\vspace{2.0em}
\section*{Methods}

\noindent\textbf{Crystal Growth and Structural Characterization}
\vspace{0.5em}

The Gallium-Indium binary flux method, as reported in \cite{CGG}, was used for the synthesis of high-quality single crystals of RGaGe (R = La, Ce). The constituent elements \textit{R}, Ga, Ge, and In were placed in an alumina crucible in a molar ratio of 1:2:1:8, and subsequently sealed in a quartz tube under vacuum. The temperature profile for the growth consisted of an initial ramp to 1050 $^\circ$C, followed by a 24 hour dwell, and a subsequent slow cooling to 500 $^\circ$C at a rate of 2.5 $^\circ$C/h. This produces well-formed platelets with a maximum dimension of 3$\cross$3$\cross$0.5 mm$^{3}$. Structural characterization was performed using a Bruker D8 Advance X-ray diffractometer with Cu K$_{\alpha}$ radiation and energy dispersive X-ray spectroscopy (EDS). 

\vspace{1.0em}
\noindent\textbf{Magnetic and Thermodynamic Measurements}
\vspace{0.5em}

Magnetization measurements were carried out in a QD Magnetic Property Measurement System (MPMS) using a vibrating sample magnetometer (VSM) option. For low-temperature magnetization measurements down to 0.4 K, a $^{3}$He probe was used with the DC option. Thermodynamic properties were measured on a Quantum Design (QD) Dynacool physical property measurement system (PPMS)-14 T, employing the two-$\tau$ relaxation method for specific heat and a standard four probe geometry for electrical transport measurements. Low-temperature specific heat measurements down to 50 mK were conducted using the dilution refrigerator (DR) option in PPMS.

\vspace{1.0em}
\noindent\textbf{Electrical Transport}
\vspace{0.5em}

Electrical transport measurements were carried out on a QD Dynacool PPMS-14 T, employing a standard four probe geometry. $\rho_{xx}$ and $\rho_{yx}$ were measured under an a.c. current of 10 mA at 9.1 Hz. $\rho_{xx}$ and $\rho_{yx}$ were obtained via symmetrization and anti-symmetrization, [$\rho_{yx}$(H)$\pm\rho_{yx}$(-H)]/2 to eliminate any transverse and longitudinal contributions from voltage contacts misalignment, respectively. MR(\%) is defined as [$\rho_{xx}$(H)-$\rho_{xx}$(0)]/$\rho_{xx}$(0)$\cross$100, where $\rho_{xx}$(H) and $\rho_{xx}$(0) are the resistivities measured at magnetic field H and zero field, respectively.

\vspace{1.0em}
\noindent\textbf{Density-functional Theory Calculations}
\vspace{0.5em}

Density functional theory (DFT) calculations for CeGaGe were performed by using the code of the Vienna ab-initio simulation package \cite{DFT_1} with the experimental lattice parameters and atomic positions as input. We used the pseudo-potential projector augmented-wave method with an energy cutoff of 350 eV for the plane-wave basis set \cite{DFT_2}. Exchange-correlated effects were treated using Perdew-Burke-Ernzerhof (PBE) generalized gradient approximation density functional \cite{DFT_3, DFT_4}. For the localized 4$f$ electrons in the paramagnetic state, the 4$f$ electrons were treated as core states, and the corresponding 4$f$ open-core bands were computed. To calculate the anomalous Hall conductivity (AHC), Bloch wavefunctions were initially projected onto localized atomic-like orbitals using the LOCPROJ implementation in VASP. These projections were used to generate the maximally localized Wannier functions (MLWFs). The resulting tight-binding Hamiltonian was subsequently used in WannierTools \cite{WT} to interpolate the electronic band structure and calculate AHC.

\section*{Acknowledgments}

The experimental work at Rice University has been primarily supported by the Vannevar Bush Faculty Fellowship ONR-VB N00014-24-1-2048 (A, HB, EM) and the Robert A. Welch Foundation Grant No. C-2114 (KA, SM, EM). We also acknowledge partial support from the US DOE BES DE-SC0026179 (EM, KR) and Rice Creative Ventures (GH,EM). The theory work from Rice has been primarily supported by the NSF Grant No.\ DMR-2220603 (YF, KSL), the Robert A. Welch Foundation Grant No.\ C-1411 (MM) and  the Vannevar Bush Faculty Fellowship ONR-VB N00014-23-1-2870 (MM, QS), with partial support provided by the DOE, BES Grant No.\ DE-SC0026179 (QS). K.-S.L. acknowledges the Carl and Lillian Illig Postdoctoral Fellowship from the Smalley-Curl Institute at Rice University. The computational calculations have in part been performed on the Shared University Grid at Rice funded by NSF under Grant EIA-0216467, a partnership between Rice University, Sun Microsystems, and Sigma Solutions, Inc., the Big-Data Private-Cloud Research Cyberinfrastructure MRI-award funded by NSF under Grant No. CNS-1338099, and the Extreme Science and Engineering Discovery Environment (XSEDE) by NSF under Grant No. DMR170109. This research used resources of the National Energy Research Scientific Computing Center (NERSC), a DOE Office of Science User Facility supported by the Office of Science of the U.S. Department of Energy under contract no. DE-AC02-05CH11231 using NERSC award BES-ERCAP0020966. The work at Los Alamos National Laboratory (LANL) was carried out under the auspices of the U.S. Department of Energy (DOE) National Nuclear Security Administration under Contract No. 89233218CNA000001, and was supported by the LANL Laboratory Directed Research and Development Program, and in part by the Center for Integrated Nanotechnologies, an Office of Science User Facility operated by the DOE Office of Science, in partnership with the LANL Institutional Computing Program for computational resources.

\clearpage

\title{Chiral Weyl-Kondo semimetallic state through enhanced correlation in CeGaGe}
\date{\today}
\label{ExtractParameter}
\maketitle

\section{Structural Characterization}\label{XRD}

To confirm the crystal structure and phase purity, powder X-ray diffraction measurements were performed at room temperature, with the pattern shown in \figref{FigS1:xrd}. Neutron powder diffraction suggests a structural transition between T = 100 K and 300 K, from tetragonal I4$_{1}$md (109) to tetragonal P4$_{3}$ (78) \cite{structural_tarnsition_CeGaGe} which is also observed through our resistivity measurement near T = 230 K. Because XRD data was collected at room temperature, Rietveld refinement was performed using only room-temperature structure I4$_{1}$md. All observed peaks (red open circles) are accounted for by the calculated peaks (black solid lines) plus the Bragg reflections (blue vertical markers) for the I4$_{1}$md space group with lattice parameters: a = b = 4.2883 \AA, c = 14.5757 \AA, in agreement with the previous report \cite{CGG}.

\setcounter{figure}{0}
\renewcommand{\thefigure}{S\arabic{figure}}
\begin{figure}[h]
\includegraphics[width=1.0\columnwidth,origin=b]{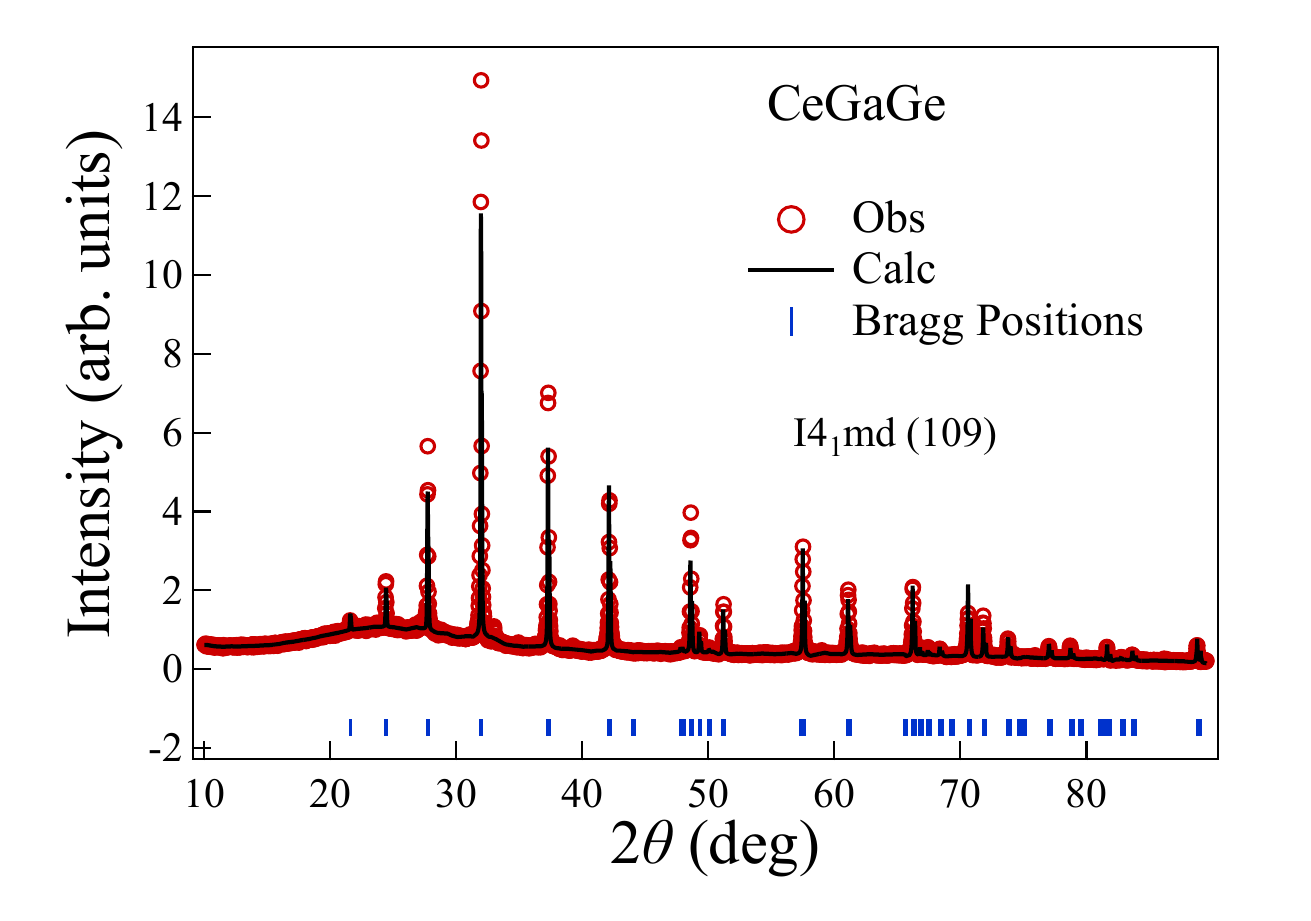}
\caption{\label{FigS1:xrd} Powder XRD pattern of CeGaGe and the Rietveld refinement using I4$_{1}$md space group are shown by red open circles and black solid line, respectively. Small blue bars show the Bragg reflection positions.}
\end{figure}

\renewcommand{\thefigure}{S\arabic{figure}}
\begin{figure}[t!]
\includegraphics[width=1.0\columnwidth,origin=b]{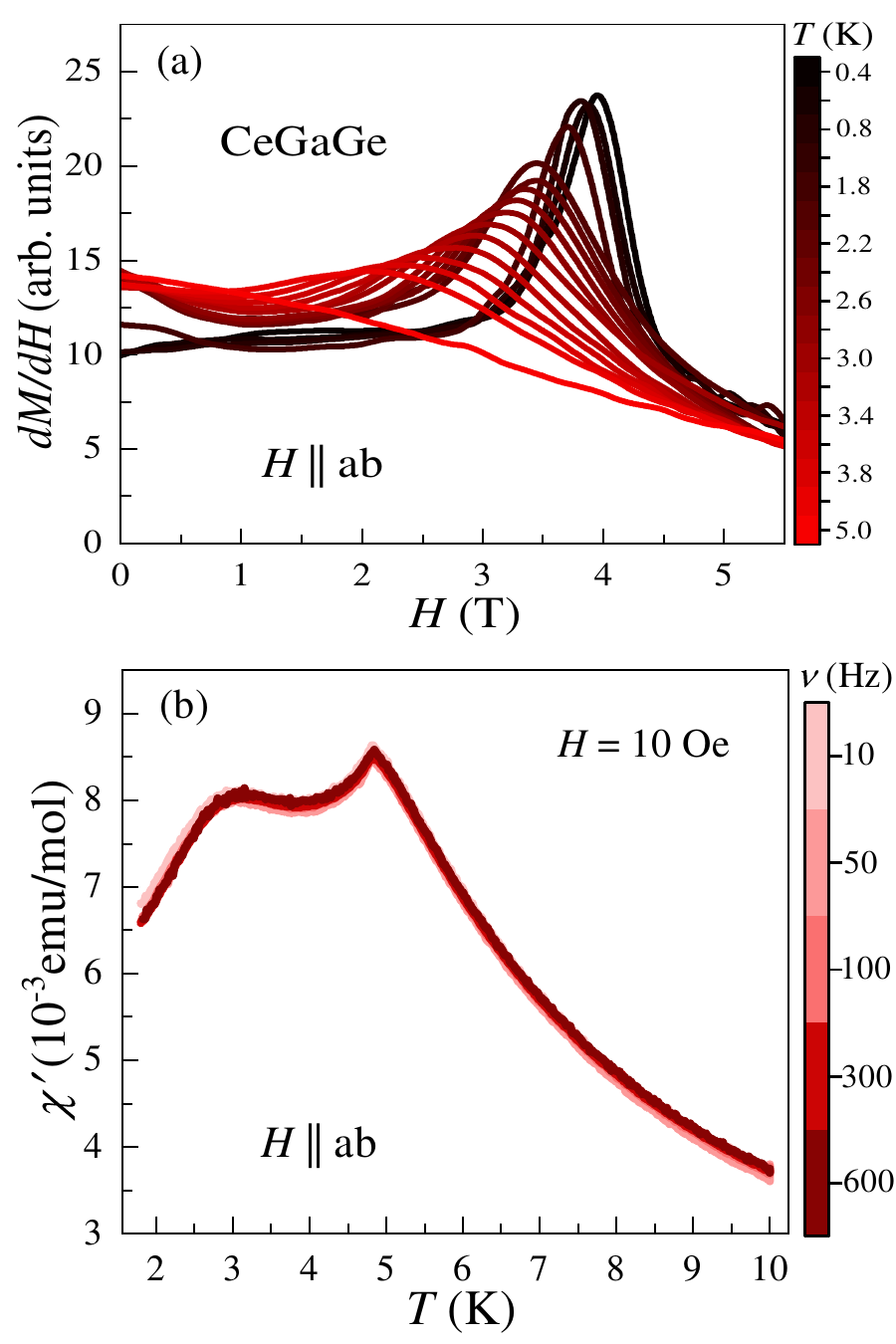}
\caption{\label{FigS2:MH_AC} a) Field-dependence of derivative of magnetization from 0.4 K - 5 K for H || $ab$. b) Temperature dependence of the real part of the AC susceptibility from 10 Hz to 600 Hz with an external field of 10 Oe for H || $ab$.}
\end{figure}

\section{Magnetization}\label{Mag}

To construct the H-T phase diagram for external field H || $ab$, the derivatives dM/dH of M(H) isotherms were used (\figref{FigS2:MH_AC}a), with a sharp peak corresponding to the metamagnetic transition. AC susceptibility $\chi'$(T) measurements were performed with a fixed external field $H$ = 10 Oe and AC amplitude of 5 Oe with varying frequencies from f = 10 Hz to 600 Hz, as shown in \figref{FigS2:MH_AC}b for $H$ || $ab$. $\chi'$ exhibits a sharp peak at the ordering temperature and another broader peak around 3.5 K, in agreement with the DC susceptibility measurements shown in the main text. No frequency dependence is observed in the $\chi'$(T) data, excluding the possibility of glassy behavior.

\section{Electrical Transport}\label{ET}

Figure \ref{FigS3:rho}a demonstrates the Kondo correlations as described in the main text. Hall data collected at 100 K for H || $c$ and current in the $ab$ plane exhibit a linear behavior, excluding the two-band contribution mentioned in the main text. A linear fit (\figref{FigS3:rho}b) shown by the red line provides an estimate of the Hall coefficient R$_{H}$ = 1.85$\cross$10$^{-9}$ $\Omega$ m/T. By using R$_{H}$ = 1/$ne$ where $n$ is the carrier density, and $ e$ is the electronic charge, we obtained $n$ = 3.36$\times10^{27} m^{-3}$.

\begin{figure}[h]
\includegraphics[width=1.0\columnwidth,origin=b]{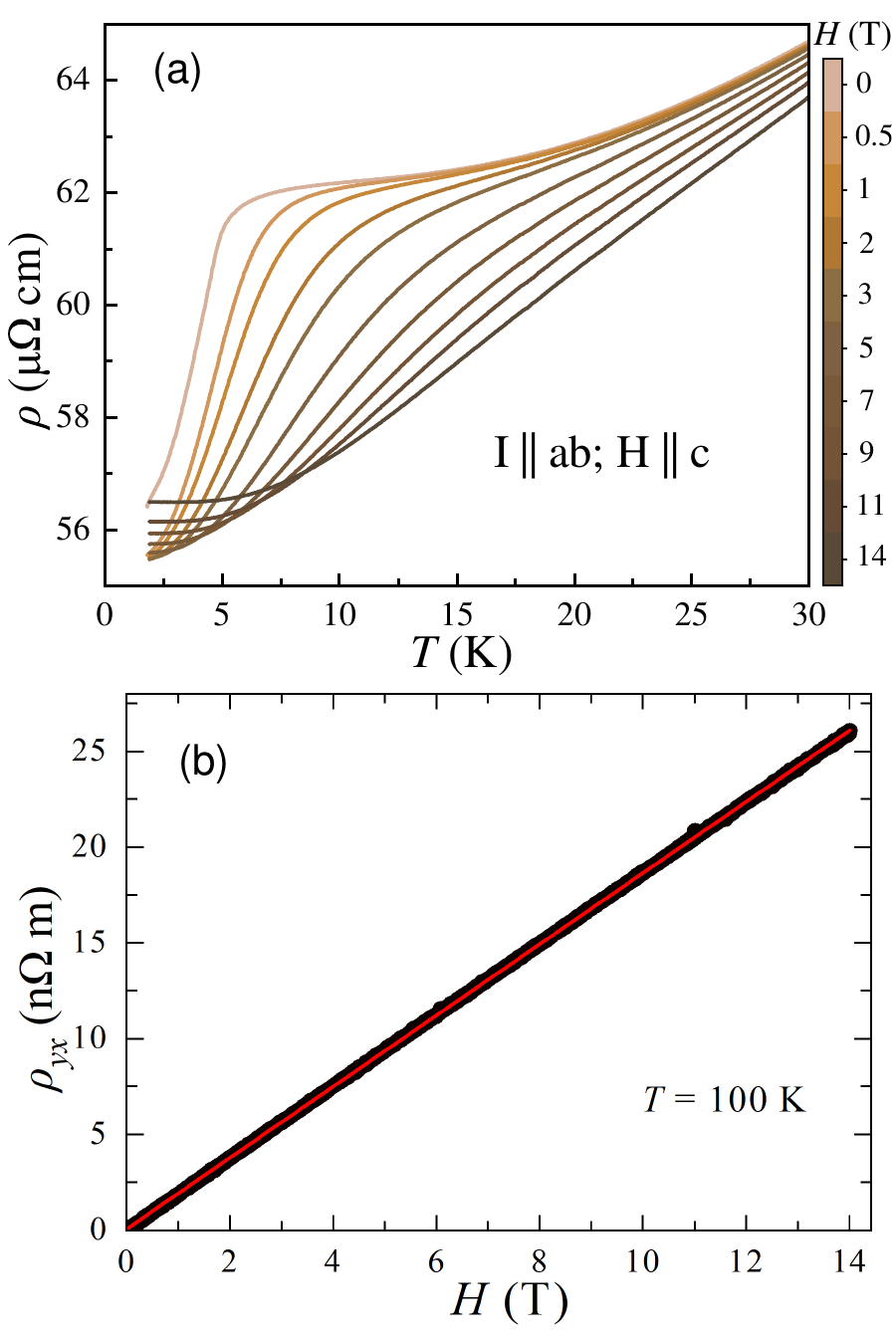}
\caption{\label{FigS3:rho} a) Low-temperature electrical resistivity for current in the $ab$ plane and H || $c$, at applied fields up to 14 T to show the presence of Kondo correlations. b) Field dependence of Hall resistivity at 100 K with a linear fit represented by a red line.}
\end{figure}

\section{Specific Heat}\label{SH}

A fit of C$_{p}$/T = $\gamma$+$\beta$T$^{2}$ where $\gamma$, and $\beta$ are the electronic, and phononic contributions, respectively to LaGaGe dataset yields $\gamma$ = 1.74 mJ/mol-K$^{2}$, $\beta$ = 0.31 mJ/mol-K$^{4}$, confirming negligible electronic correlations in the non-magnetic LaGaGe compound. From the high temperature fit of C$_p$/T above $T_\mathrm{N}$ with a short temperature interval (11 - 20 K), a $\gamma$ $\approx$ 45 mJ/mol-K$^{2}$, and $\beta$ = 0.41 mJ/mol-K$^{4}$ are obtained.

\begin{figure}[h]
\includegraphics[width=1.0\columnwidth,origin=b]{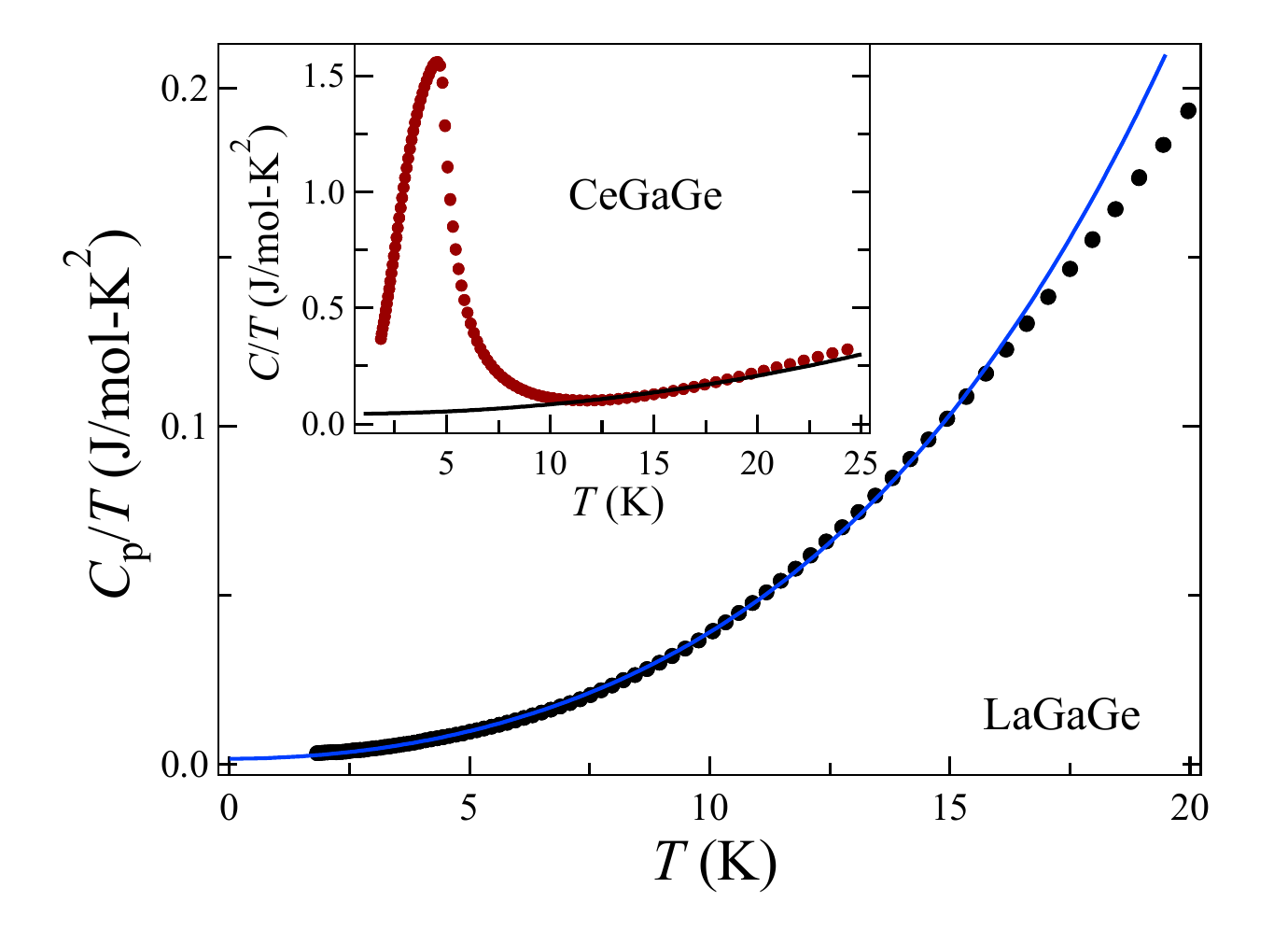}
\caption{\label{FigS4:SH} Temperature dependence of C$_{p}$/T for non-4$f$ analog LaGaGe at zero applied fields, where the blue line represents the fit, C$_{p}$/T = $\gamma$ + $\beta$T$^{2}$. The inset shows the $\gamma$ + $\beta$T$^{2}$ fit (black line) for CeGaGe above the ordering temperature.}
\end{figure}

\begin{figure}[b]
\includegraphics[width=1.0\columnwidth,origin=b]{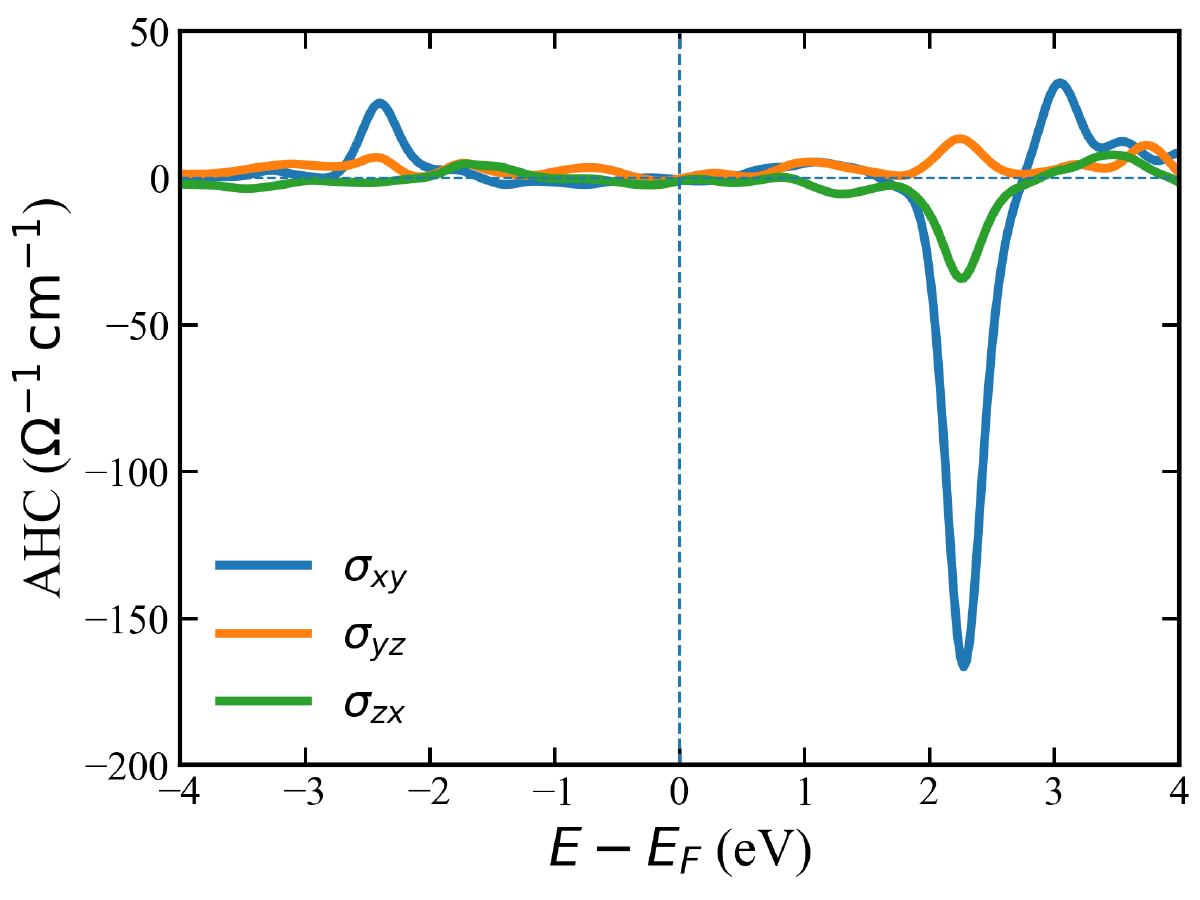}
\caption{\label{FigS5:AHC} Energy dependence of calculated AHC representing $\sigma_{xy}$ (blue), $\sigma_{yz}$ (orange), and $\sigma_{zx}$ (green).}
\end{figure}

\section{Anomalous Hall Conductivity}\label{AHC_wo_f}

The anomalous Hall conductivity (AHC) was obtained from a Wannier interpolated tight-binding Hamiltonian constructed from localized orbital projections and subsequently evaluated using WannierTools. The symmetry analysis predicts symmetry-enforced crossings that can contribute to the Berry curvature and hence to the AHC. The most pronounced feature to the AHC is found in $\sigma_{xy}$, consistent with the anomalous Hall response measured experimentally in the $xy$ channel. However, it occurs away from the Fermi level, suggesting the relevant symmetry-enforced crossings might be located away from the Fermi level. At the same time, the precise position of the Fermi level in CeGaGe can differ from the DFT value because of small deviations in stoichiometry, defects or other sample dependent effects. Such a shift in chemical potential could therefore bring the experimental Fermi level closer to the calculated AHC. Important observation is that the absolute value of calculated AHC is smaller than the experimentally observed value. Since the 4$f$ states were treated as core states in this calculation, the reduced magnitude suggests that the 4$f$-electron states may provide an important additional contribution to the experimentally observed anomalous conductivity response.

\section{Symmetry Classification}\label{SC}

The crystal structure of CeGaGe investigated in this study has the tetragonal space group no.~78 ($P 4_3$), which is chiral, noncentrosymmetric, and nonsymmorphic.
In the following, we denote the space group operation by $\{ R | n_1, n_2, n_3 \}$, where $R$ is a point group operation, and $n_1$, $n_2$, $n_3$ collectively denote a translation of $n_1 \mathbf{a}_1 + n_2 \mathbf{a}_2 + n_3 \mathbf{a}_3$ with $\mathbf{a}_1$, $\mathbf{a}_2$, $\mathbf{a}_3$ being the Bravais lattice vectors.
In other words, upon the $\{ R | n_1, n_2, n_3 \}$ operation, a position $\mathbf{x}$ is transformed into $R \mathbf{x} + n_1 \mathbf{a}_1 + n_2 \mathbf{a}_2 + n_3 \mathbf{a}_3$.
Space group no.~78 ($P 4_3$) contains nonsymmorphic screw rotations $\{ C_{4z} | 0,0,3/4 \}$ and $\{ C_{2z} | 0,0,1/2 \}$, where $C_{4z}$ and $C_{2z}$ denote the $\pi/2$ and $\pi$ rotations, respectively, along the Cartesian $z$ axis.
Due to the absence of any symmetry with its point group being mirror reflection, spatial inversion, or roto-inversion, space group no.~78 ($P 4_3$) is chiral. In addition, the lack of spatial inversion symmetry renders the space group no.~78 ($P 4_3$) noncentrosymmetric.
In the paramagnetic phase that preserves the spinful time-reversal $\mathcal{T}$ symmetry, a system with space group no.~78 ($P 4_3$) could host multiple band-crossing features \cite{symmetry_enforced, topology_chiral_crystal_Weakly}:
\begin{enumerate}
    \item Weyl points located along the high-symmetry lines $\Gamma-Z$ and $M-A$ from the accordion-type band connectivity enforced by the $\{ C_{4z} | 0,0,3/4 \}$ symmetry.
    \item Weyl points located along the high-symmetry line $X-R$ from the hourglass-type band connectivity enforced by the $\{ C_{2z} | 0,0,1/2 \}$ symmetry.
    \item Weyl nodal plane on the $k_z = \pi$ boundary of the Brillouin zone enforced by the composite symmetry $\{ C_{2z} \mathcal{T} | 0,0,1/2 \}$.
    \item Kramers Weyl points at time-reversal-invariant momenta due to the chiral crystal structure.
\end{enumerate}
The abundance of Weyl band crossings provides a natural resource of Berry curvature. In the presence of an external magnetic field, Weyl band crossings can be gapped or deformed.

\section{Anomalous Hall effect analysis}

Here we discuss the intrinsic anomalous Hall effect (AHE) induced by the Berry curvature~\cite{higher_order_non-linear_hall_effect} for the CeGaGe structure investigated in this study. In particular, symmetries can constrain the lowest-order AHE. In the paramagnetic phase, CeGaGe has the magnetic space group $P 4_3 1'$, which hosts a second-order AHE due to its magnetic point group $4 1'$. In the presence of an external magnetic field $H \parallel c$, CeGaGe has the magnetic space group $P 4_3$, which hosts a first-order AHE due to its magnetic point group $4$. In the presence of an external magnetic field $H \parallel ab$, CeGaGe has the magnetic space group $P 1$, which hosts a first-order AHE due to its magnetic point group $1$.

\section{Toy model study of $\rm CeGaGe$} 
We construct a toy model with two elementary band representations (EBRs), each representing two orbitals at specific Wyckoff positions, to mimic the band structure and orbital characters near the Fermi energy obtained from DFT.
The orbitals are at $4a$ Wyckoff position with coordinates
$\mathbf{r}_A=(000)$, $\mathbf{r}_B=(00\frac14)$, $\mathbf{r}_C=(00\frac12)$, $\mathbf{r}_D=(00\frac34)$ in units of lattice constants.
We denote the two EBRs (orbitals) $c_1$ and $c_2$.
The orbitals form a basis $\Psi^\dagger= (\Psi^\dagger_1, \Psi^\dagger_2)$ and 
$\Psi^\dagger_I = (c^\dagger_{I A \uparrow},c^\dagger_{I A \downarrow}, \epsilon c^\dagger_{I B \uparrow}, \bar{\epsilon}c^\dagger_{I B \downarrow}, \epsilon^2 c^\dagger_{I C \uparrow}, \bar{\epsilon}^2 c^\dagger_{I C \downarrow}, \epsilon^3 c^\dagger_{I D \uparrow}, \bar{\epsilon}^3 c^\dagger_{I D \downarrow})$, 
where $\epsilon=e^{i3\pi/4}$, $\bar{\epsilon}=e^{-i3\pi/4}$, $I=1,2$ is the EBR index.
Such a basis is generated with coset decomposition
\begin{equation}
    SG/\mathbb{T} = E G_q + g G_q + g^2 G_q + g^3 G_q
\end{equation}
where $SG$ is the space group $P4_{3}$, $g=\{C_{4z}^3|0,0,\frac14\}$ is the screw symmetry, $G_q=\{E\}$ is the site-symmetry group of $q=\mathbf{r}_A$ in the $4a$ Wyckoff position and $\mathbb T$ is the lattice translation (abelian normal) subgroup.
From this basis, we can systematically determine the symmetry-allowed independent hoppings. \cite{disconnected_band_representation, symmetry_indicators, cWKSM_1}

The conduction electron noninteracting Hamiltonian that preserves crystalline and time-reversal symmetries takes the form \cite{cWKSM_78_1}
\begin{align} \label{eqn:H0}
    H_0 = \Psi^\dagger h_0 \Psi \,, \quad
    h_0 = \begin{pmatrix}
        h_{11} & h_{12} \\ 
        h_{12}^\dagger & h_{22}
    \end{pmatrix}
\end{align}
We consider nearest-neighbor hopping and spin-orbit coupling (SOC). Then each block matrix is given by:

\begin{align}
h_{II}(\mathbf{k}) &=
\begin{aligned}[t]
&\begin{pmatrix}
0 & A_{0} & 0 & A_{0}^{\dagger} \\
A_{0}^{\dagger} & 0 & -A_{0} & 0 \\
0 & -A_{0}^{\dagger} & 0 & A_{0} \\
A_{0} & 0 & A_{0}^{\dagger} & 0
\end{pmatrix} \\
&+
\begin{pmatrix}
B_{0} & 0 & 0 & 0 \\
0 & B_{1} & 0 & 0 \\
0 & 0 & B_{2} & 0 \\
0 & 0 & 0 & B_{3}
\end{pmatrix}
+2p_{9}\cos k_z\,\mathbf{1}_8
\end{aligned}
\\[1ex]
h_{12}(\mathbf{k}) &=
\begin{pmatrix}
0 & A_{1} & 0 & A_{2}^{\dagger} \\
A_{2}^{\dagger} & 0 & -A_{1} & 0 \\
0 & -A_{2}^{\dagger} & 0 & A_{1} \\
A_{1} & 0 & A_{2}^{\dagger} & 0
\end{pmatrix}.
\end{align}
where each submatrix takes the form:
\begin{align}
c_z &= \cos\frac{k_z}{4}, & \nonumber
s_z &= \sin\frac{k_z}{4},\\ \nonumber
c_x &= \cos k_x, &
s_x &= \sin k_x,\\ \nonumber
c_y &= \cos k_y, &
s_y &= \sin k_y,\\ \nonumber
\alpha_\pm &= p_1 \pm i p_2,\qquad &
\beta_\pm &= p_3 \pm i p_4.
\end{align}

\begin{align}
    A_{0} &= 
\begin{pmatrix}
\alpha_+c_z+(-i \alpha_+)s_z &
\beta_+c_z+(-i \beta_+)s_z\\
-\beta_-c_z+(i \beta_-)s_z &
\alpha_-c_z-(i \alpha_-)s_z
\end{pmatrix} \\
    A_{i} &= 
    \begin{pmatrix}
        p_{Ai} c_z - i p_{Ai} s_z & 0 \\
        0 & p_{Ai} c_z - i p_{Ai} s_z
    \end{pmatrix} \,,\quad i=1,2 \\ 
    B_{0} &= \begin{pmatrix}
        2 p_{5} c_x - 2 p_{6} s_x & 2 \left(i p_{7} - p_{8}\right) s_x \\
        - 2 \left(i p_{7} + p_{8}\right) s_x & 2 p_{5} c_x + 2 p_{6} s_x
    \end{pmatrix} \\ 
    B_{1} &= \begin{pmatrix}
        2 p_{5} c_y + 2 p_{6} s_y & 2 \left(- i p_{7} + p_{8}\right) s_y \\
        2 \left(i p_{7} + p_{8}\right) s_y & 2 p_{5} c_y - 2 p_{6} s_y
    \end{pmatrix} \\ 
    B_{2} &= \begin{pmatrix}
        2 p_{5} c_x + 2 p_{6} s_x & 2 \left(- i p_{7} + p_{8}\right) s_x \\
        2 \left(i p_{7} + p_{8}\right) s_x & 2 p_{5} c_x - 2 p_{6} s_x
    \end{pmatrix} \\ 
    B_{3} &= \begin{pmatrix}
        2 p_{5} c_y - 2 p_{6} s_y & 2 \left(i p_{7} - p_{8}\right)s_y \\
        - 2 \left(i p_{7} + p_{8}\right) s_y & 2 p_{5} c_y + 2 p_{6} s_y
    \end{pmatrix}
\end{align}
Here $p_a=p_a^{IJ}$ is the hopping parameter, and the EBR indices $I,J$ are omitted in the expression for simplicity. 
In addition, $A_0$ is the out-of-plane nearest-neighbor intra-orbital hopping term and $B_0$, $B_1$, $B_2$, $B_3$ are the in-plane nearest-neighbor intra-orbital hopping terms (with SOC included), whereas $A_1$ and $A_2$ are the out-of-plane nearest-neighbor inter-orbital hopping terms. The parameters we used are summarized in Table~\ref{tab:params}.

\begin{table}[]
    \centering
    \begin{tabular}{c|rrrrrrrrr}
    \toprule 
        $IJ$ &$p_1$ &$p_2$ &$p_3$ &$p_4$ &$p_5$ &$p_6$ &$p_7$ &$p_8$ &$p_9$ \\
        \hline 
        $11$ &$-1.0$ &$-0.5$ &$-0.2$ &$-0.2$ &$2.0$ &$0.5$ &$-0.2$ &$-0.2$ &$-1.0$\\ 
        $22$ &$1.0$ &$0.5$ &$0.2$ &$0.2$ &$-1.0$ &$0.0$ &$0.2$ &$0.2$ &$1.0$\\ 
    \midrule 
        $IJ$ &$p_{A1}$ &$p_{A2}$  & & & & \\
        \hline
        $12$ &$1.0$ &$0.5$ & & & \\ 
    \midrule 
        &$\epsilon_f$ &$V_1$  &$V_2$ &$\nu$ &&$r$ &$\lambda$ &$\mu$ \\
        \hline
        &$-1.00$ &$0.80$ &$0.80$ &$3$ &&$0.191$ &$1.20$ &$0.199$\\ 
    \bottomrule
    \end{tabular}
    \caption{Parameter values of the toy model and the saddle point solution results. The unit is the nearest neighbor hopping amplitude $t$, i.e., $|p_1|=t$. }
    \label{tab:params}
\end{table}

The band structure of this toy model is shown in the main text.
We choose parameters so that the band structure mimics the DFT results at low energies.

We further add a localized $f$ orbital with Hubbard interaction to this model, forming a periodic Anderson lattice model. The $f$ orbitals couple to the two conduction electrons through a Kondo hybridization term. The Hamiltonian is
\begin{align}
    H = H_0 + \sum_{i,\sigma} \left[ \epsilon_f n_{f,i\sigma} + \sum_{I=1,2} V_I \left( f_{i\sigma}^\dagger c_{I,i\sigma} + h.c. \right) \right] \nonumber\\  + \sum_i U n_{f,i\uparrow} n_{f,i\downarrow} -\mu \sum_{i,\sigma} \left( n_{c_1,i\sigma}+n_{c_2,i\sigma}+n_{f,i\sigma} \right) \,, 
\end{align}

where $i$ is the site label, representing both the unit cell and sublattice labels, and $f_{i\sigma}^\dagger$ is the creation operator for a $f$ electron at site $i$ with spin $\sigma$. 
Here $\mu$ is the chemical potential and $n_{c_I,i\sigma}=c^\dagger_{I,i\sigma}c_{I,i\sigma}$ is the particle number operator for orbital $I$ at site $i$ with spin $\sigma$, $n_{f,i\sigma}=f_{i\sigma}^\dagger f_{i\sigma}$ is the $f$ electron number operator.
In the infinite-$U$ limit, the number of $f$ electrons on each site is constrained by $\sum_\sigma \langle n_{f,i\sigma}\rangle \leq 1$.
Hereafter, $\langle \cdots \rangle$ is understood as the ground state expectation value at zero temperature or the thermal average at finite temperature. We analyze the interaction effects based on an auxiliary boson saddle point approach \cite{Kondo_problem_to_heavy_fermion_physics}. 
To treat the constraint, an auxiliary boson operator $b$ and a Lagrange multiplier $\lambda$ are introduced such that the periodic Anderson model in the infinite-$U$ limit is represented as follows:
\begin{align} \label{eqn:Hab} 
    H = H_0 + \sum_{i,\sigma} \left[ \epsilon_f n_{f,i\sigma}  + \sum_{I=1,2} V_I \left( (b_i f_{i\sigma}^\dagger) c_{I,i\sigma} + h.c. \right) \right] \nonumber\\ + \lambda \sum_{i} \left( \sum_{\sigma} n_{f,i\sigma} + b^\dagger_i b_i -1 \right) \nonumber\\ 
    -\mu \sum_{i,\sigma} \left( n_{c_1,i\sigma}+n_{c_2,i\sigma}+n_{f,i\sigma} \right) \
\end{align}

\begin{figure}[h]
\includegraphics[width=1.0\columnwidth,origin=b]{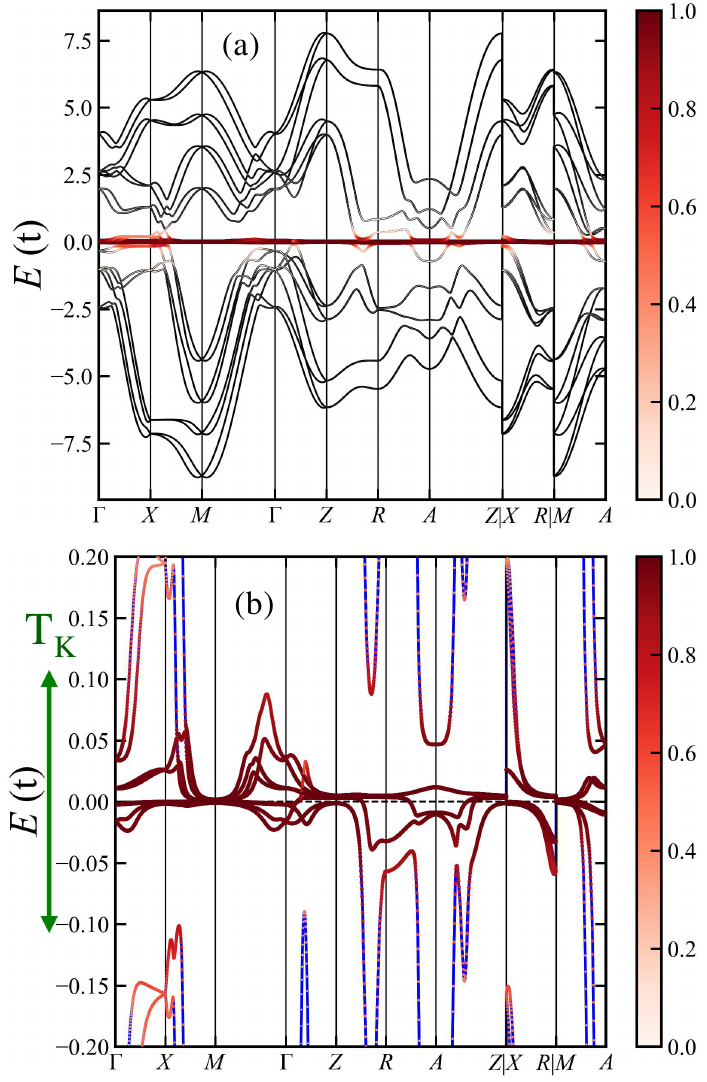}
\caption{\label{FigS5:band_struc} (a) Saddle point solution. (b) Zoom-in view of the heavy fermion bands. Red color indicates the $f$-electron component of each band.}
\end{figure}

At the saddle-point level, a solution can be obtained by considering the condensation of the auxiliary boson operator, which means effectively replacing $b_i$ by $\langle b \rangle$. We may absorb the complex phase of $\langle b \rangle$ by the $f$ electron operators such that we further replace $\langle b \rangle$ by $r$, where $r$ is a real number.
This leads to the following self-consistent equations
\begin{align}
    & r = \sqrt{1 - \sum_{\sigma} \left\langle f^\dagger_{i\sigma} f_{i\sigma} \right\rangle} \,, \\
    & \lambda = \sum_{I=1,2} \frac{-V_I}{r} \sum_{\sigma} \mathrm{Re} \left\langle c^\dagger_{I,i\sigma} f_{i\sigma} \right\rangle \,.
\end{align}
Here we have taken the expectation values $\left\langle f^\dagger_{i\sigma} f_{i\sigma} \right\rangle$ and $\left\langle c^\dagger_{I,i\sigma} f_{i\sigma} \right\rangle $ for each sublattice to be identical due to symmetries. 
In addition, the chemical potential is determined by fixing the total particle number throughout the calculations:
\begin{equation}
    \sum_{i\sigma}\langle n_{c_1,i\sigma} + n_{c_2,i\sigma} +  n_{f,i\sigma} \rangle = \nu N_L \,, 
\end{equation}
where $N_L$ is the total number of sites and $\nu$ is the electron filling.
We set $\nu=3$, which ensures each site is half-filled.
These equations can be solved self-consistently to obtain the saddle-point solution. The bare parameters $\epsilon_f$, $V_1$, $V_2$ and the self-consistent parameters $r$, $\lambda$, $\mu$ are listed in Table~\ref{tab:params}. The resulting bandwidth of $f$-band defines a Kondo temperature $T_{\mathrm{K}} \approx 0.2t$.

According to the symmetry analysis discussed in the previous section, $\cal T$ symmetry must be broken to get nonzero AHE.
In our experimental setup, $\cal T$ is broken by magnetization in the system. Here, we consider the paramagnetic phase with a small external magnetic field.
When the magnetic moment is magnetized by this external magnetic field, we consider the following phenomenological mean-field Hamiltonian 
\begin{align}
    H' = H + m_1 \sum_{iIss'} M \cdot \sigma_{ss'} c^\dagger_{iIs}c_{iIs'} \nonumber\\
    + m_2 \sum_{iss'} M \cdot \sigma_{ss'} f^\dagger_{is}f_{is'}
\end{align}
where $M=M(B,T)$ is the ordered moment, and the Zeeman terms break time-reversal symmetry. Here, $m_1$ and $m_2$ are the effective Zeeman couplings for $c$ and $f$ electrons. The effective Zeeman term $m_1$ mostly comes from Kondo coupling, while $m_2$ has two origins: RKKY interaction and magnetization-induced effective magnetic field. During the calculations, we set $m_1 = m_2 = m$. It should be noted that the conventional Zeeman term due to the external magnetic field has a much lower energy scale in the small magnetic field regime. 

\renewcommand{\thefigure}{S\arabic{figure}}
\begin{figure}[h]
\includegraphics[width=1.0\columnwidth,origin=b]{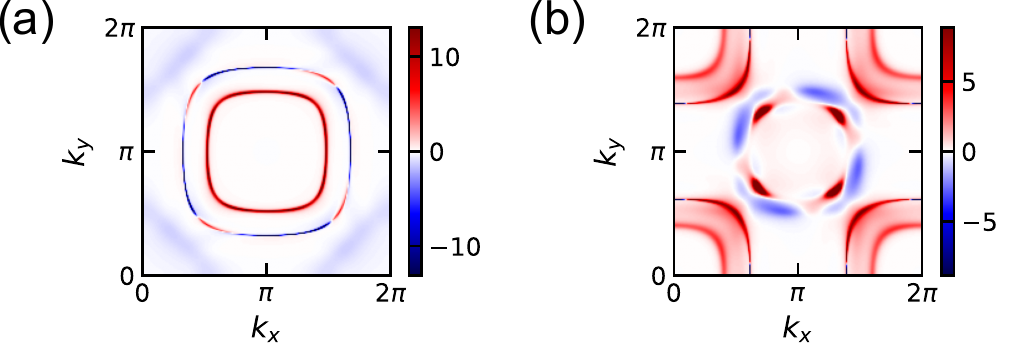}
    \caption{2D Berry curvature plots for (a) $k_z=0$ and (b) $k_z=\pi$.}
    \label{FigS4:bcmap}
\end{figure}

The intrinsic AHE is proportional to the integral of filled states' $U(1)$ Berry curvature.
We compute the Berry curvature of filled bands by 
\begin{equation}
    \Omega_{ij}(\mathbf {k}) = \sum_{n,m} \frac{\langle n| \frac{\partial H}{\partial k_i}|m\rangle \langle m |\frac{\partial H}{\partial k_j}|n\rangle }{(\epsilon_n - \epsilon_m)^2} (f_n - f_m) \,, 
\end{equation}
where $f_n$ is the Fermi-Dirac distribution for band $n$ at momentum $\mathbf k$. 
Then the AHE is given by 
\begin{align}
    \sigma_{ij}^{A} &= -\frac{e^2}{\hbar} \int \frac{d\mathbf k}{(2\pi)^3} \Omega_{ij}(\mathbf {k}) \\ 
    &= -\frac{e^2}{\hbar} \frac{|\mathbf{a}_i\times \mathbf{a}_j|}{V_{\rm u.c.}} Q_{ij}
\end{align}
where $V_{\rm u.c.}=|\mathbf{a}_1\times \mathbf{a}_2\cdot  \mathbf{a}_3|$ is the volume of unit cell, $Q_{ij} = \int \frac{d\widetilde{\mathbf {k}}}{(2\pi)^3} \Omega_{ij}(\widetilde{\mathbf {k}})$ is the dimensionless Berry curvature integral and $\widetilde{\mathbf k}$ is the dimensionless reciprocal lattice vector.

In our model, a large AHE response could come from the avoided crossings and the deformed crossings of the topological nodes due to the Zeeman terms, where the former include the nodal planes in the $k_z=\pi$ plane, and the latter include the Kramers Weyl points at time-reversal invariant momenta and symmetry enforced Weyl points along high symmetry lines and also accidental Weyl points. The 2D Berry curvature maps of the bottom 12 bands at $k_z=0$ and $k_z=\pi$ are shown in \figref{FigS4:bcmap}(a,b). In this calculation, we use a grid size $200\times 200$ and fix $mM=0.01t\approx 0.05T_{\mathrm{K}}$. In each plot, the color bar is clipped at 99\% of the maximum value.

We calculated the intrinsic anomalous Hall conductance $\sigma$ using our toy model as a proof of principle, demonstrating that the Kondo effect amplifies the response. We choose the lattice constant $a_3=14.5414$ \AA~ and lattice size $60\times 60 \times 60$. Clearly, the $f$ electron dominated heavy fermion bands have a giant contribution to AHE as shown in the main text because (i) they are pinned to the Fermi energy and (ii) they give rise to enhanced responses to external drives due to the narrow bandwidth.

\end{document}